\documentclass[twocolumn]{aastex631}

\newcommand{\Msun}{{\rm M_{\odot}}}
\newcommand{\Lsun}{{\rm L_{\odot}}}

\newcommand{\kpc}{\, {\rm kpc}}
\newcommand{\pc}{\, {\rm pc}}
\newcommand{\kmps}{\, {\rm km \, s^{-1}}}

\newcommand{\alphaCO}{{\alpha_{\rm CO}}}

\newcommand{\GDR}{{\delta_{\rm GDR}}}
\newcommand{\ICO}{{I_{\rm CO}}}
\newcommand{\LCO}{{L_{\rm CO}}}

\shorttitle{The CO-to-H$_2$ Conversion Factor in M83}
\shortauthors{A. M. Lee et al.} 

\usepackage{natbib}

\begin{document}

\title{The CO-to-H$_2$ Conversion Factor in the Barred Spiral Galaxy M83}

\correspondingauthor{Amanda M Lee}
\email{amamlee@umass.edu}

\author[0000-0001-8254-6768]{Amanda M Lee}
\affiliation{Department of Physics and Astronomy, Stony Brook University, Stony Brook, NY 11794-3800}
\affiliation{Department of Astronomy, University of Massachusetts Amherst, 710 North Pleasant Street, Amherst, MA 01003}

\author[0000-0002-8762-7863]{Jin Koda}
\affiliation{Department of Physics and Astronomy, Stony Brook University, Stony Brook, NY 11794-3800}

\author[0000-0002-0465-5421]{Akihiko Hirota}
\affiliation{NAOJ Chile, National Astronomical Observatory of Japan, Los Abedules 3085 Office 701, Vitacura, Santiago 763-0414, Chile}
\affiliation{Joint ALMA Observatory, Alonso de C\'ordova 3107, Vitacura, Santiago 763-0355, Chile}

\author[0000-0002-1639-1515]{Fumi Egusa}
\affiliation{Institute of Astronomy, Graduate School of Science, The University of Tokyo, 2-21-1 Osawa, Mitaka, Tokyo 181-0015, Japan}

\author[0000-0002-3871-010X]{Mark Heyer}
\affiliation{Department of Astronomy, University of Massachusetts Amherst, 710 North Pleasant Street, Amherst, MA 01003}

\begin{abstract}
We analyze the CO-to-H$_2$ conversion factor ($\alphaCO$) in the nearby barred spiral galaxy M83.
We present 
new HI observations from the JVLA and single-dish GBT in the disk of the galaxy, and 
combine them with 
maps of CO(1-0) integrated intensity and dust surface density from the literature.
$\alphaCO$ and the gas-to-dust ratio ($\GDR$) are simultaneously derived in annuli of 2 kpc width from R = 1-7 kpc.
We find that
$\alphaCO$ and $\GDR$ both increase radially, by a factor of $\sim$ 2-3 from the center to the outskirts of the disk. The luminosity-weighted averages over the disk are $\alphaCO = 3.14$ (2.06, 4.96) $\Msun\,\pc^{-2}[{\rm K \cdot}  \kmps]^{-1}$ and $\GDR$ = 137 (111, 182) at the 68\% (1$\sigma$) confidence level.
These are consistent with the $\alphaCO$ and $\GDR$ values measured in the Milky Way.
In addition to possible variations of $\alphaCO$ due to the radial metallicity gradient, we test the possibility of variations in $\alphaCO$ 
due to changes in the underlying cloud populations, as a function of galactic radius.
Using a truncated power-law molecular cloud CO luminosity function and an empirical power-law relation for cloud-mass and luminosity, we show that the changes in the underlying cloud population may account for a factor of $\sim 1.5-2.0$ radial change in $\alphaCO$.
\end{abstract}

\section{introduction} \label{sec:intro}
Molecular hydrogen (H$_2$) is the most abundant molecule in the interstellar medium (ISM) and is important for star formation and galaxy evolution. 
However, since the cold H$_2$ is not detectable in emission, molecular gas masses are often traced using emission from the second most abundant molecule CO.
This assumes a constant H$_2$ mass to CO luminosity ratio, deemed the CO-to-H$_2$ conversion factor \citep[$X_{{\rm{CO}}}$ or $\alphaCO$; ][and references therein]{Bolatto:2013ys}.\footnote{
Typically, the conversion factor denoted as $X_{{\rm{CO}}}$ does not include the contribution from heavier elements (e.g., helium) in mass and is expressed in units of $\rm H_2\,cm^{-2} \left[K\,\cdot \kmps\right]^{-1}$.
$\alphaCO$ includes the heavier elements, by multiplying 1.36 to the H$_2$ mass, and is in units of $\rm \Msun\,\pc^{-2}\,\left[K\,\cdot \kmps\right]^{-1}$.}

There are four main techniques used to determine the conversion factor: (1) the virial method, which assumes clouds are gravitationally-bound and virialized \citep[e.g.][]{Sanders:1984fx, Solomon:1987pr}, (2) $\gamma$-ray emission from interactions of cosmic-rays and protons (hydrogens), with an assumed model for the cosmic ray distribution \citep[e.g.][]{Lebrun:1983aa, Strong:1996fr}, (3) dust emission as a tracer of the total gas surface density, assuming that the gas-to-dust ratio ($\GDR$) is known or measurable \citep[e.g.][]{Reach:1998aa, Dame:2001gs, Planck-Collaboration:2011qy, Sandstrom:2013vn}, and (4) multi-line CO observations to trace gas densities, assuming all the line emission come from the same regions \citep{Garcia-Burillo:1993ys, Israel:2009oq, Cormier:2018aa, Israel2020, Teng2022, Teng2023, den-Brok:2023aa}.

There are also new efforts to use (5) [CI] or [CII] line emission, assuming that their flux can be converted to H$_2$ mass \citep{Accurso2017, Madden:2020aa}.
Some other methods are less frequently used and are applied only to some specific regions (e.g., solar neighborhood) including
(6) stellar extinction \citep{Lombardi:2006aa}, and (7) X-ray shadows \citep{Sofue:2016ab}.
The assumptions embedded in all these methods can be questioned, 
but they have given a roughly consistent $\alphaCO$ 
in the Milky Way (MW) and in nearby galaxies with solar metallicity, with a caveat that a factor of 2-3 uncertainty remains, even with all the measurements combined \citep{Bolatto:2013ys}.

The dust-based method (3) uses measurements of
the HI and dust surface densities, $\Sigma_{\rm HI}$ and $\Sigma_{\rm d}$,
and CO surface brightness, $I_{\rm CO}$.
$\GDR$ 
is first determined in HI-dominant regions
as $\GDR = \Sigma_{\rm HI}/\Sigma_{\rm d}$.
Assuming that this $\GDR$ is the same in H$_2$-dominant regions,
the conversion factor is calculated as $\alphaCO=(\GDR \Sigma_{\rm d} - \Sigma_{\rm HI})/\ICO$
\citep{Reach:1998aa, Dame:2001gs}.
\citet{Leroy:2011lr} and \citet{Sandstrom:2013vn} expanded this method
and fit the HI and H$_2$-dominated regions simultaneously in a large sample of nearby star-forming galaxies at kpc-scale resolutions with the fitting equation,
\begin{equation}
      \Sigma_{\rm d}
        =  \frac{1}{\GDR} (\Sigma_{\rm HI} + \alphaCO I_{\rm CO}). \label{eq:fiteq0}
\end{equation}
They used CO($J$=2-1) emission, instead of CO($J$=1-0), for $I_{\rm CO}$ from single-dish observations \citep{Leroy:2009zv}
and $\Sigma_{\rm HI}$ from interferometric data from the Very Large Array \citep{Walter:2008mw}.
The $\Sigma_{\rm HI}$ term provides the absolute mass scale in this equation, and thus sets the absolute scales of $\GDR$ and $\alphaCO$.
They found an average $\alphaCO$ marginally smaller than, but still consistent with that of the MW value, with the exception of the galaxy centers, where $\alphaCO$ had values of up to an order of magnitude lower.

In this paper, we present new HI 21-cm line emission observations in M83 with the Jansky Very Large Array (JVLA) and Green Bank Telescope (GBT) and produce a high-quality image of the atomic hydrogen distribution via their combination.
Together with CO (J=1-0) emission data from the ALMA single-dish telescopes \citep{Koda:2020aa},
as well as dust emission data from the Herschel Space Observatory \citep{Bendo:2012ab} we determine $\alphaCO$ and $\GDR$ in M83 by simultaneously fitting for them.

M83 is a close nearby barred spiral galaxy located at a distance of 4.5 Mpc \citep{Thim:2003aa}. It has been the subject of many studies for its similarities to the Milky Way and near face-on inclination of 26 $^\circ$ \citep{Koda:2023aa}. 

\section{HI Data}

New HI 21-cm data of M83 were taken with the JVLA and GBT.
These data were reduced and imaged along with archival JVLA data.

\subsection{Jansky Very Large Array (JVLA)}\label{sec:jvla}

\subsubsection{New and Archival Data}

We used a total of 30 measurement sets (MSs) from the JVLA,
including 22 archival MSs from projects 13B-194 and 14B-192 (PI: Bigiel).
The archival data mosaicked 10 adjacent fields around M83,
and we used only the data of the central pointing on the optical disk of M83.
The additional 8 new MSs from project 21B-113 (PI: Koda) observed the single central pointing
with higher sensitivity at long baselines. 
Each MS represents an observing track starting from observations of
a bandpass calibrator, and then cycling between the gain calibrator and target fields.
At the frequency of the HI 21-cm line ($\nu_{\rm HI}$ = 1420.405751 MHz, $\lambda_{\rm HI}=$ 21.1061141~cm),
the primary beam of the JVLA antennae has a full-width at half power (FWHP) of 29.6'.

\begin{figure*}[h]
\centering
    \includegraphics[width=0.5\textwidth]{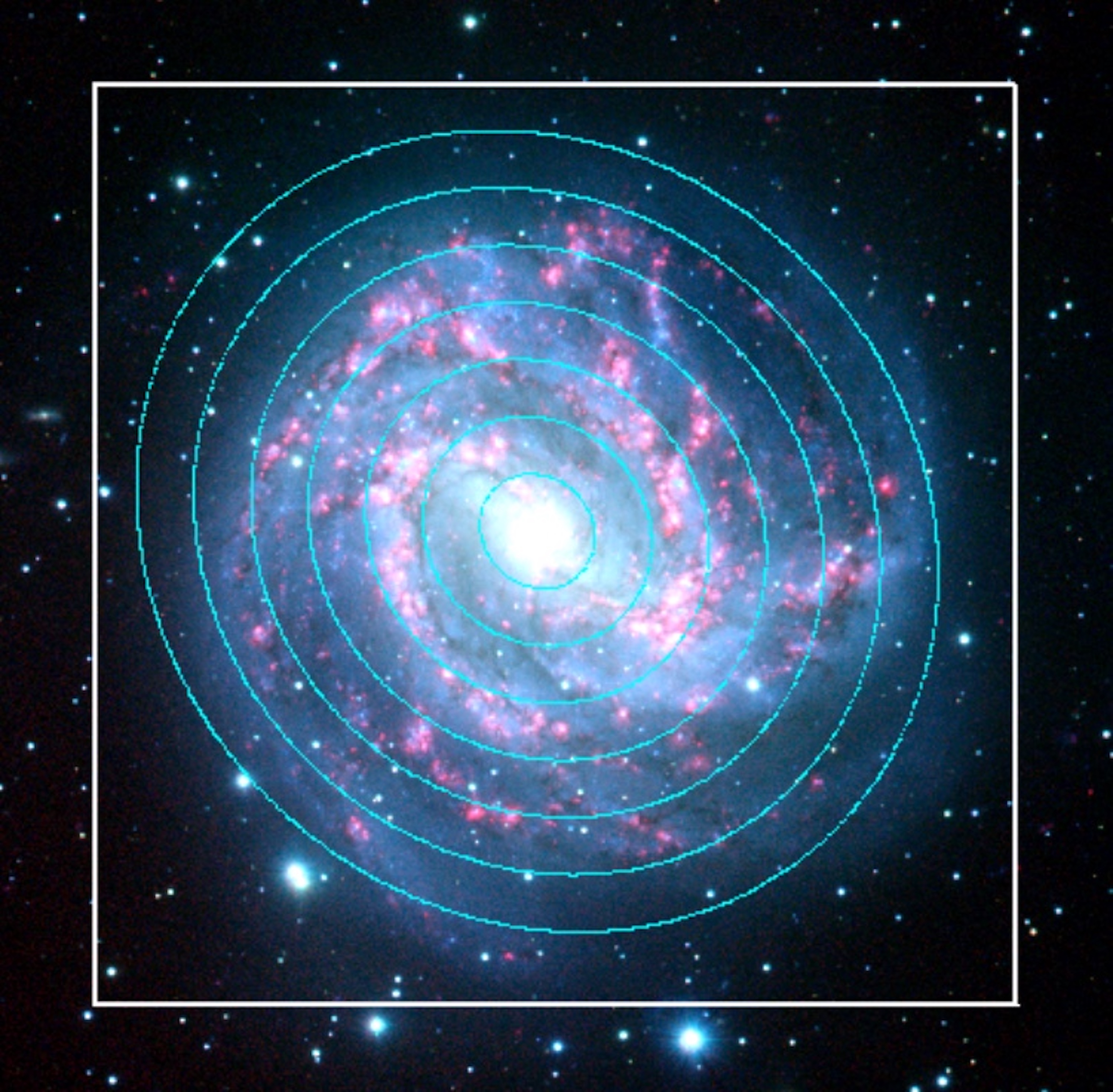}
\caption{An optical B-V-H$\alpha$ composite image of M83 taken at the Cerro Tololo Inter-American Observatory and downloaded from the NASA/IPAC Extragalactic Database \citep{Meurer2006, Cook:2014ab}. The white box indicates the $11.7\arcmin\times11.7\arcmin$ area observed in CO(1-0) \citep{Koda:2020aa}.
The coverage in HI is significantly larger than this area, with the VLA primary beam size of $29.6\arcmin$ (FWHM) and GBT coverage even larger.
Radial contours (cyan) show the radii from $R$=1-7~kpc at a 1~kpc interval with the position angle and inclination of $225\deg$ and $26\deg$, respectively \citep{Koda:2023aa}.
}
\label{fig:fov}
\end{figure*}

\subsubsection{Data Reduction} \label{sec:jvlared}

The data were reduced using version 5.6.2-2 of
the Common Astronomy Software Applications package \citep[CASA; ][]{McMullin:2007aa, Bean:2022aa}.
The MSs include 10 spectral windows (SPWs).
We used only one narrow band SPW with the HI line and one broad-band SPW for calibration.
The archival MSs have a narrow-band SPW with 2048 channels with a channel width of 1.95 kHz ($\sim$ 0.4 km/s) and the new MSs have 4096 channels with a 0.97 kHz ($\sim$ 0.2 km/s) width.

The standard calibration strategy was applied as described in the CASA guide\footnote{
\url{https://casaguides.nrao.edu/index.php/VLA_high_frequency_Spectral_Line_tutorial_-_IRC\%2B10216-CASA5.5.0}
},
including flagging for ``shadow", ``clip", and ``quack",
and corrections for antenna positions, elevation-dependent gain variations, and opacity.
The flux scale was derived using the flux of 3C286, our bandpass calibrator, measured by the observatory.
Delay and bandpass solutions were applied to the entire track.
We then derived time-variant gain solutions with a broad-band SPW,
and applied them to both the broad and narrow-band SPWs. 
Some bad data with unrealistically high amplitudes or phase scatters
became obvious after these calibrations.
We flagged those in the raw data and repeated all the calibrations from scratch.

For imaging we used only the narrow-band SPW.
We subtracted the continuum emission in the \textit{uv} space using line-free channels,
and regridded each MS to a common velocity grid, covering the velocity range
of 200-800$\kmps$ with a channel width of 5$\kmps$ for HI.
Imaging was performed with the CASA task ``tclean".
The final JVLA data cube has a beam size of $9\arcsec.89 \times 8\arcsec.77$
and a sensitivity of 0.76 mJy/beam in each channel at the field center.

\subsection{Green Bank Telescope (GBT)} \label{sec:gbt}
\subsubsection{Observations} \label{sec:gbtobs}

GBT observations were also made (project 20A-432).
We used the single-element L-band dual-polarization receiver
and the VErsatile GBT Astronomical Spectrometer (VEGAS).
We employed one of the eight banks in the narrowest bandwidth, 11.72~MHz with 32768 channels.
The corresponding velocity coverage is about 2,400~$\kmps$ at a 0.08~$\kmps$ channel width.
The beam size at the full width half maximum (FWHM) is $\sim8.7\arcmin$ at the HI frequency/wavelength.

A total of 5.25 hours was allocated over two sessions in May 2020.
Each session started from receiver tuning, and telescope pointing and focus corrections using 1311-2216.
A region of about $1.2\arcdeg \times 1.5\arcdeg$ around M83 was mapped
with the On-The-Fly (OTF) mapping technique in the RA and DEC directions,
with a step size of $3.6\arcmin$ ($=\frac{1}{2} \lambda_{\rm HI}/D$, between OTF scans, where $D=100\,\rm m$ is the dish diameter).
Each OTF mapping in RA and DEC required 26 and 22 scans to cover the whole region,
and each scan took 72 and 90 seconds, respectively.
After every 4-5 scans, we integrated an OFF position of (RA,DEC)=(13:30:00.0,-29:51:19.0) for 30 seconds.
We obtained a total of 6 OTF maps (3 in RA and 3 in DEC).
We observed 3C286 for flux calibration.

\subsubsection{Data Reduction} \label{sec:gbtred}

The data were reduced in a standard way with the GBTIDL software and pipeline.
Time and elevation-dependent amplitude variations were calibrated
using the noise diode in the telescope, ON/OFF sky measurements, and $T_{\rm sys}$ measurements.
The calibrated spectra were gridded spatially to produce a data cube using
the GaussBessel kernel function.
We adopted spatial and velocity pixel scales of $1.47\arcmin$ and 0.75 km/s. 

In order to correct for relative flux variations,
we made 12 data cubes for the two polarizations of the 6 OTF maps.
We measured their flux ratios and inverted the ratio matrix to derive relative flux scaling factors.
Their range was small and was 0.98-1.02 (peak-to-peak) after setting the median to be unity.
We applied these factors to the data and made the integrated data cube.

The absolute flux scale was calibrated using 3C286.
This calibrator had been monitored and confirmed to be stable \citep{Perley:2013aa}
with the flux density of $S_{\nu}$ = 14.97~Jy at 1.420~GHz \citep{Perley:2017aa}.

\subsection{Combination of JVLA and GBT Data} \label{sec:joint}

The JVLA and GBT cubes were combined with the ``feather" task in CASA.
We regridded the GBT cube, applied the JVLA primary beam to it, combined it with the JVLA cube,
and applied the JVLA primary beam correction to the combined cube.
The final JVLA+GBT data cube has a beam size of $9\arcsec.89 \times 8\arcsec.77$ and
a sensitivity of about 1.1 mJy/beam in each $5\kmps$ channel at the field center. 
This HI data covers a large area, and in this paper,
we focus only on the disk of M83 (Figure \ref{fig:fov}).

We applied a mask to the final cube for generating an integrated intensity map for a 11'$\times$11' box around the galaxy center (the region analyzed in Section \ref{sec:HIcomp}; see Figure \ref{fig:archivedata}).
We made the mask in two steps.
First, we utilized the clean component cube from the JVLA imaging (Section 2.1.2), smoothed it with a Gaussian kernel with a FWHM of $60\arcsec$ (about the width of the HI spiral arms), and took regions with  fluxes $>5\,\rm mJy/beam$ in the smoothed cube to make the first mask with pixel values of either 0 or 1. The first mask encloses significant emission in the JVLA imaging.
Second, we expanded this mask by smoothing it with a Gaussian kernel with a FWHM of $90\arcsec$ and by taking the regions  $>0.1$.
We start from the clean component cube because, by definition, it includes only the flux components identified as emission.
This operation is to include any extended emission from the GBT data.
The parameters in these operations were chosen manually by looking at the emission distributions in the cubes, and conservatively, so that the final mask may include blank sky but does not miss any emission.
We used this mask also to calculate a noise map using the standard noise propagation based on the number of velocity pixels in each spatial pixel.
Within the 11'$\times$11' region, we calculated the total flux with and without applying the mask in the cube, and found that the ratio is $\sim$ 1.09. Thus, the mask likely encloses all of the emission.

\subsection{Comparisons with Previous HI Data}\label{sec:HIcomp}
\begin{figure*}[h]
    \centering
    \includegraphics[width=1.0\textwidth]{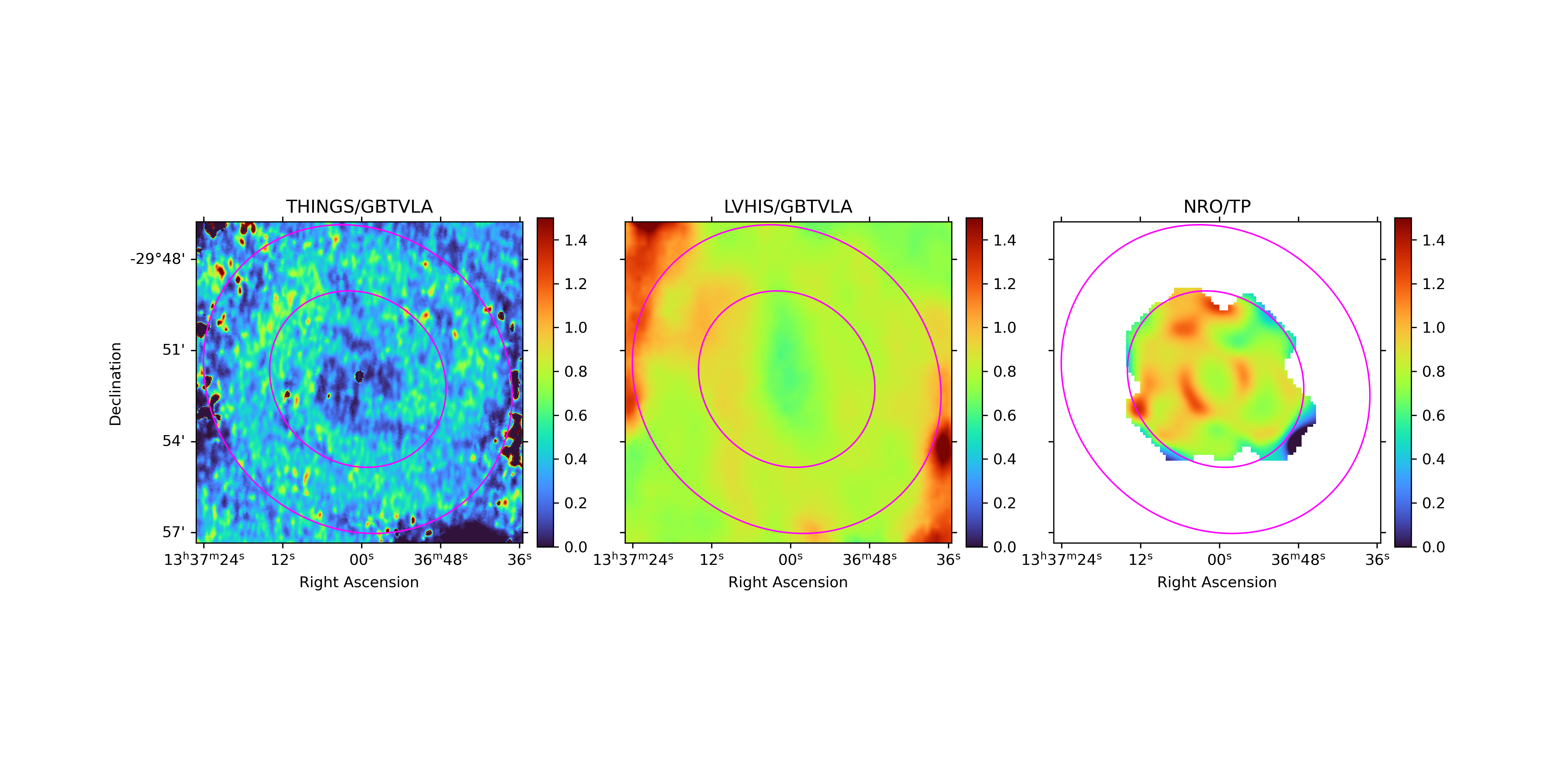} 
    \caption{Ratio maps between archival HI and CO(1-0) data, and the data used in this study.
    \textit{Left:} THINGS/GBTVLA. Middle: LVHIS/GBTVLA.
    \textit{Right:} NRO/TP. GBTVLA and TP denote the HI and CO(1-0) data used in this study, while the archival HI and CO(1-0) data are denoted by THINGS and LVHIS, and NRO respectively. The map resolutions are at the lower resolution between the two used in the ratio (see text). Radial contours at R = 4 and R = 7 kpc are overlaid in magenta. In the 11'$\times$11' region for the left two panels (HI data), the ratio of the total fluxes are THINGS/GBTVLA $\sim$ 0.39 and LVHIS/GBTVLA $\sim$ 0.82 (see text). For the rightmost panel (CO data), the intensity-weighted average ratio is $\sim$ 0.83 within $R\leq4\kpc$.}
    \label{fig:archivedata}
 \end{figure*}

To check for consistency with previous HI data, we acquired archival HI data cubes and integrated intensity (moment 0) maps of M83 from the THINGS survey \citep[The HI Nearby Galaxy Survey][]{Walter:2008mw},
which achieved a high resolution of $15\arcsec.16\times 11\arcsec.44$ with natural weighting without single-dish data,
and the LVHIS survey \citep[The Local Volume HI Survey; ][]{Koribalski:2018vj}, which achieved a lower spatial resolution of $\sim 85\arcsec.8 \times 60\arcsec.8$ but with single-dish data for flux recovery. 
While the THINGS survey also distributes data at a resolution of $10\arcsec.40 \times 5\arcsec.60$ with uniform weighting without single-dish data, we used the naturally weighted data for comparison, since it typically reproduces the emission distribution more accurately.
We make two comparisons: (1) between their archival data cubes and our cube to evaluate the quality of the data and (2) between their distributed moment 0 maps and our map to understand the difference in derived $\alphaCO$ (section \ref{sec:comparisons}). For (1) and (2) we smoothed and regridded our data to the resolution and pixel scale of the archival data. For the flux comparisons, we consider the ratio of total fluxes in the 11'$\times$11' region shown in Figure \ref{fig:archivedata}.

For the first comparison, the ratio between the total flux of the archival data cubes and ours, without applying the mask, is THINGS/GBTVLA $\sim$ 0.39 and LVHIS/GBTVLA $\sim$ 0.92, where GBTVLA denotes the data from this work. After applying the mask, the ratios are THINGS/GBTVLA $\sim$ 0.65 and LVHIS/GBTVLA $\sim$ 0.84. Because of negatives in the THINGS data, the THINGS/GBTVLA ratio with and without the mask vary, and depend on the mask.

For the second comparison, we compare the archival moment 0 maps to our map because in Section \ref{sec:comparisons}, we will compare our results with those from previous studies, which used the distributed moment 0 map 
\citep{Foyle:2012aa, Jiao:2021aa, Yasuda:2023aa}.
These moment 0 maps were made with some masks applied, although we do not have access to those masks.
The ratio between the total fluxes within the 11'$\times$11' region is THINGS/GBTVLA $\sim$ 0.39 and LVHIS/GBTVLA $\sim$ 0.82. Figure \ref{fig:archivedata} shows the ratio maps. The THINGS data show the galactic structures in HI (e.g., spiral arms) clearly at the high resolution, but suffer from missing flux, due to the lack of single-dish data. 
The fraction of recovered flux varies systematically as a function of the galactic structures. 
For example, it is as high as $\sim$80\% along the spiral arms, but decreases to $\sim$20\% in the interarm regions. This systematic variation is expected since interferometric data without single-dish data tend to miss extended emission.
On the other hand, the LVHIS data is relatively more consistent with our new map, though it has a much lower resolution. The LVHIS HI fluxes also appear consistent with recent HI observations in M83 \citep[][see their Figure 3]{Eibensteiner:2023aa}.

We see that the THINGS/GBTVLA ratio varies significantly depending on what mask was applied/not applied. Overall, the THINGS flux appears lower than that of our new data and analysis.

\section{CO(1-0) and Infrared Data}

 We compare the distributions of atomic hydrogen (in HI, discussed above), molecular hydrogen (in CO), and dust (in far-infrared).

We adopt the CO(J=1-0) integrated intensity map (and noise map)  from \citet{Koda:2020aa}.
This was obtained via observations with ALMA's Total Power (TP) array, which consists of four single-dish telescopes. 
The map has a low spatial resolution of $56.6\arcsec$ (FWHM; $\sim$ 1.2 kpc).
We use the integrated intensity map in the main beam temperature in units of $\rm K\cdot \kmps$.
As a consistency check,
we compare this map to the CO(1-0) map obtained from the Nobeyama 45-m telescope \citep{Kuno:2007zr} in Figure \ref{fig:archivedata}. 
The Nobeyama 45-m data is higher in resolution ($15\arcsec$) but is limited in spatial coverage (radial extent $R \sim 4$ kpc). 
The intensity weighted average ratio between the Nobeyama CO(1-0) and ALMA TP maps within $R=$ 0-4~kpc is $\sim$ 0.8.

As a tracer of the dust distribution,
we use the Herschel SPIRE 500$\rm{\mu m}$ image \citep{Bendo:2012ab, Foyle:2013aa}. The final calibration uncertainties in the dust map include a systematic uncertainty of
$\sim$5\% and 
random uncertainty of $\sim$2\% \citep{Bendo:2012ab}.

The image is obtained from the NASA/IPAC data archive
for the Very Nearby Galaxy Survey (VNGS) \citep{VNGS}
\footnote{\url{https://irsa.ipac.caltech.edu/data/Herschel/VNGS/images/M83/}}.

\begin{figure*}[h]
    \centering
    \includegraphics[width=1.0\textwidth]{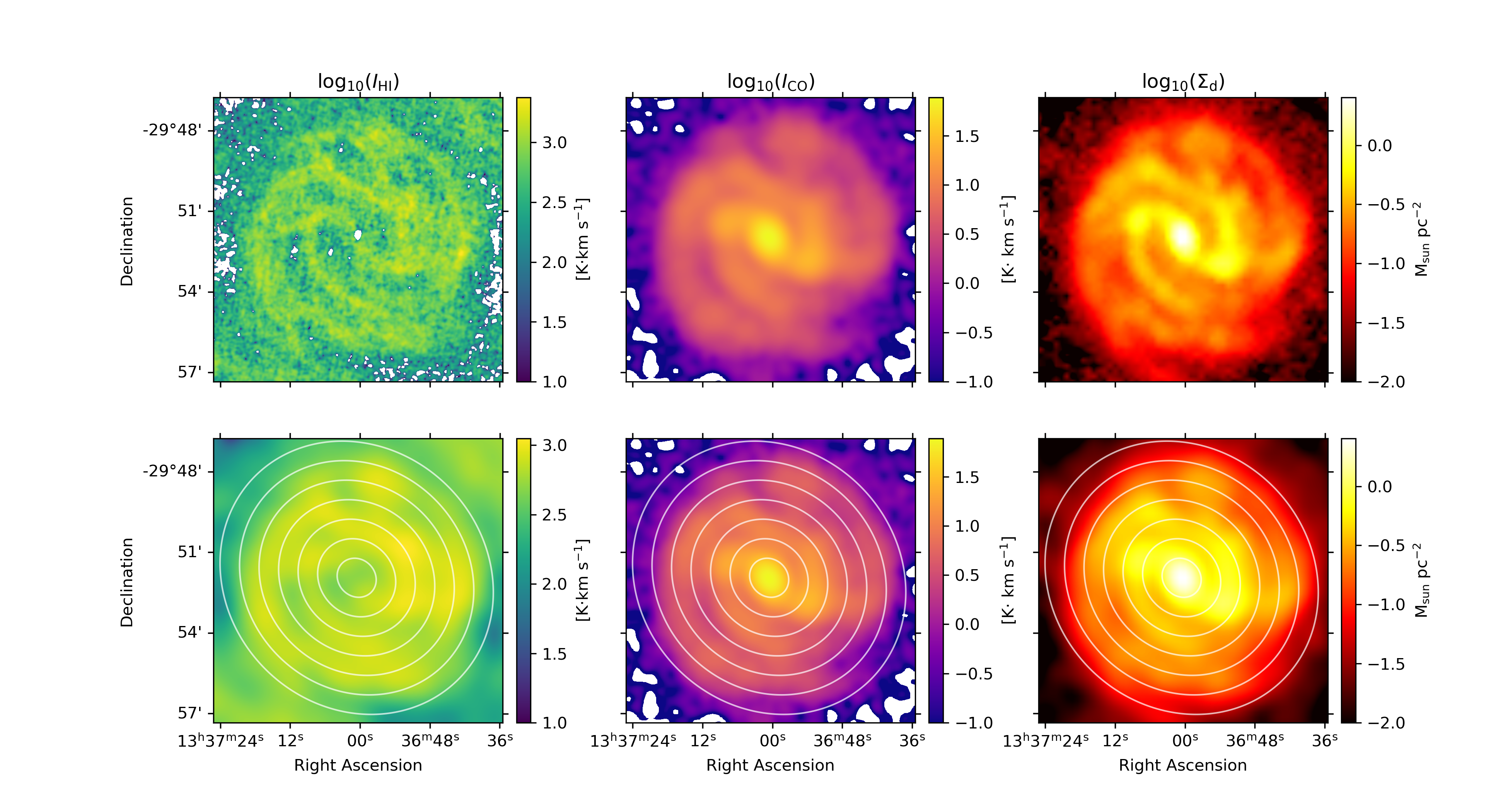}
    \caption{The $I_{{\rm HI}}$, $I_{{\rm CO}}$, and $\Sigma_{{\rm {d}}}$ maps from left to right, shown at their original resolutions (top) and smoothed to the resolution of the CO(1-0) map (56.6$\arcsec$) (bottom). Radial contours from R=1-7 kpc are overlaid in white. We exclude the central $R <$ 1~kpc in the analysis, since it suffers from HI absorption.}
    \label{fig:IcoHImapsv2}
\end{figure*}

\section{Methods}
We derive the CO-to-H$_2$ conversion factor ($\alphaCO$)
and gas-to-dust mass ratio ($\GDR$)
from the observations of $\Sigma_{\rm d}$, $\Sigma_{\rm HI}$, and $\ICO$
using eq. (\ref{eq:fiteq0}).
$\Sigma_{\rm HI}$ and the molecular gas surface density $\Sigma_{\rm H_2}$,
include the masses of heavier elements (e.g., helium).
Figure \ref{fig:IcoHImapsv2} shows the HI, CO, and far-infrared intensity images that we use for the analysis.
Here we describe how to evaluate $\Sigma_{\rm HI}$, $\Sigma_{\rm H_2}$, and $\Sigma_{\rm d}$ from the observed intensities,
and how to use them for fitting to derive $\alphaCO$ and $\GDR$.

\subsection{HI and H$_2$ Surface Densities}

$\Sigma_{\rm H_2}$ is proportional to the CO integrated intensity $I_{\rm CO}$ in units of ${\rm K} \cdot \kmps$,
\begin{equation}
    \Sigma_{\rm H_2} = \alphaCO I_{\rm CO}. \label{eq:Ico}
\end{equation}
The coefficient, $\alphaCO$, is the conversion factor and will be determined by fitting.
Following the definition by \citet{Sandstrom:2013vn}, $\alphaCO$ includes the mass of helium (a factor of 1.36)
and is in units of $\Msun\,\pc^{-2} [{\rm K} \cdot\kmps]^{-1}$.\footnote{Hereafter, we will drop the units of $\alphaCO$ and $\alpha_{\rm HI}$ for simplicity. Unless otherwise specified, their units are always $\Msun\,\pc^{-2} [{\rm K} \cdot \, \kmps]^{-1}$ and include the contribution of helium and heavier elements (a factor of 1.36).}

$\Sigma_{\rm HI}$ is derived from the HI 21cm integrated intensity $I_{\rm HI}$  in units of ${\rm K} \cdot \kmps$,
\begin{equation}
    \Sigma_{\rm HI} = \alpha_{\rm HI} I_{\rm HI},
    \label{eq:aHI}
\end{equation}
where $\alpha_{\rm HI}= 1.98\times10^{-2} \Msun\,\pc^{-2} [{\rm K} \cdot  \kmps]^{-1}$ (including the factor of 1.36 to account for helium)
is derived theoretically in an optically-thin condition under local thermodynamic equilibrium (LTE).

\subsection{Dust Surface Density}\label{sec:dust500}

\citet{Aniano:2020aa} used the \citet{Draine:2007fj} dust model, and fitted it to SEDs of a large set of nearby galaxies over a wide wavelength range of dust emission from 3.6 to 500$\mu m$.
They derived $\Sigma_{\rm d}$ maps of the galaxies,
as well as maps of other parameters, such as the dust fraction in polycyclic aromatic hydrocarbons (PAHs),
background stellar light intensity, and infrared luminosity from the dust.
So far, their study 
provides the 
most accurate $\Sigma_{\rm d}$ maps. \citet{Sandstrom:2013vn} adopted their maps for the analysis of $\alphaCO$.

M83 is not among the sample of \citet{Aniano:2020aa}.
However, \citet{Aniano:2020aa} compared their $\Sigma_{\rm d}$ maps with 500$\rm{\mu m}$ luminosity surface density maps ($\Sigma_{\nu L_{\nu}}$) and found the relation,
\begin{equation}
    \log \left( \frac{\Sigma_{\rm d}}{\Msun \, \kpc^{-2}} \right) = -0.42 + 0.942 \log \left( \frac{\Sigma_{\nu L_{\nu}}}{\Lsun\,\kpc^{-2}} \right).
    \label{eq:Aniano}
\end{equation}
They concluded that $\Sigma_{\nu L_{\nu}}$ reproduces
their $\Sigma_{\rm d}$ maps to an accuracy of $\sim$ 0.07 dex over almost three decades of $\Sigma_{\nu L_{\nu}}$ and we accordingly adopt an uncertainty of 20\% for $\Sigma_{\rm d}$. 
Since this is larger than the uncertainties in the $500\mu$m data mentioned in Bendo et al. 2012, we assume that 20\% is the overall uncertainty. 
We also assume this includes the uncertainty from sky fluctuation including background sources, since even in the outermost region considered in our analysis, most pixels (98\%) have intensities at least a factor of 5 higher than the typical sky fluctuation around M83.

They found that their $\Sigma_{\rm d}$ can be estimated more reliably with this equation, than with the more widely-used SED-fitting with an optically thin modified blackbody approximation.
A similar relation is obtained with galaxy-integrated values by \citet{Groves:2015aa}.
Hence, we adopt this equation and convert the Herschel 500$\rm{\mu m}$ image into a $\Sigma_{\rm d}$ map.

The $\Sigma_{\rm d}$ map inherits the resolution and pixel scale of the Herschel 500$\rm{\mu m}$ image, which are $39.0\arcsec$ and $12.0 \arcsec$.
We smooth this image with a Gaussian to the CO(1-0) resolution of $56.6\arcsec$. The gas-to-dust ratio (GDR) in \citet{Aniano:2020aa} is $\sim 140$ in the conditions of the local interstellar medium (ISM).
Thus, we measure the GDR relative to this value in the following sections.
The \citet{Aniano:2020aa} GDR is from the mass of atoms depleted from the gas phase in the local ISM:
the hydrogen-to-dust mass ratio (including PAHs) is 
$\sim$ 103 \citep[=1/0.0097; ][see their Table 2]{Draine:2021aa},
and by taking into account the mass of helium (a factor 1.36), their GDR is $\sim$ 140 (Bruce Draine, in private communication).

\subsection{Finding Solutions} \label{sec:findsolutions}

We search for the solutions of eq. (\ref{eq:fiteq0}) within annuli of $2 \kpc$ width
centered on
the galactic center from $R=1$ to $7 \kpc$ in $1 \kpc$ increments 
(Section \ref{sec:resultsradial}),
as well as for the whole disk area of $R=$1-7~kpc (Section \ref{sec:resultsaverage}). 
The central R $<$ 1 kpc is excluded in the analysis.
The annuli are defined on the disk plane with the position angle of $225\deg$ and inclination of $26\deg$ \citep{Koda:2023aa}.
There is some redundancy between adjacent annuli.
We adopt two methods to find the solutions.

The first method (the $\chi^2$ method) is to minimize the $\chi^2$ of the equation
(referred to as ``plane fits" in the $\Sigma_{\rm HI}$-$I_{\rm CO}$-$\Sigma_{\rm d}$ space by \citealt{Leroy:2011lr}).
The second method (the $\Delta \log \GDR$ method) is to minimize $\Delta \log \GDR$, the scatter in $\log \GDR$.
\citet{Leroy:2011lr} and \citet{Sandstrom:2013vn} tested both methods and found no significant differences in results.
They favored the $\Delta \log \GDR$ method as they found it more stable in fitting. 
Here, we use both methods with a slight preference to the $\chi^2$ method,
as it gives a convenient estimation of errors, assuming that all errors in the analysis follow the normal distribution.

\begin{figure*}[h]
    \centering
    \includegraphics[width=1.0\textwidth]{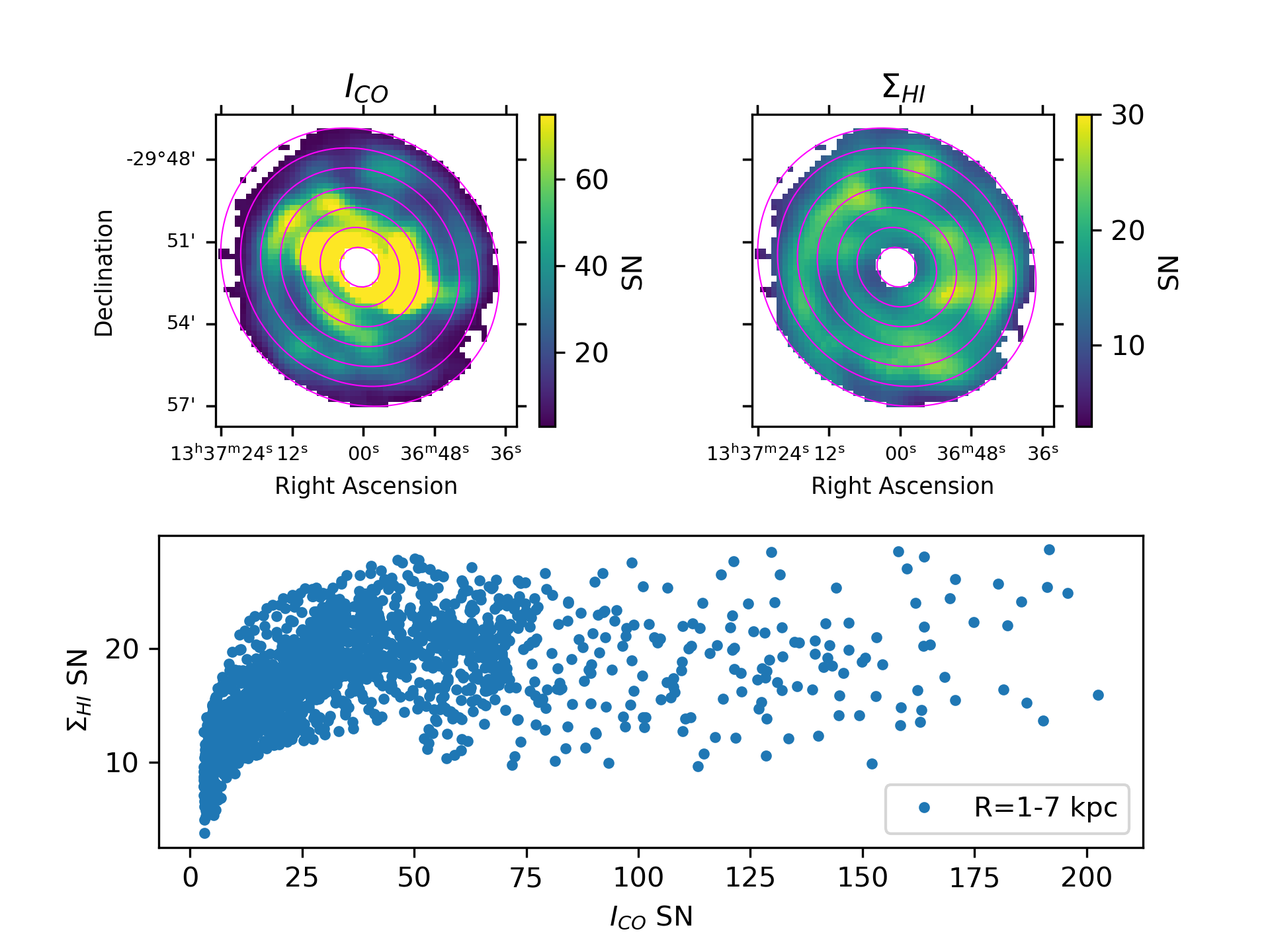}
    \caption{Comparison of the CO and HI signal-to-noise (S/N) ratios.
    \textit{Top:} The S/N maps 
    of $I_{\rm CO}$ (left) and $\Sigma_{\rm HI}$ (right), with radial contours $R=1-7$ kpc overplotted in magenta. Only the pixels used in the analyses (i.e., S/N$>3$) are shown.
    We do not show a map in $\Sigma_{\rm d}$ since we set it constant (S/N=5, see Section \ref{sec:dust500}) everywhere. 
    Note the $I_{\rm CO}$ S/N color map does not show its full range of S/N.
    \textit{Bottom:} Pixel-to-pixel correlation between the $\Sigma_{\rm HI}$ and $I_{\rm CO}$ S/N.}
    \label{fig:SNplot}
\end{figure*}

We use only the pixels with a signal-to-noise (S/N) ratio above 3 in all of the maps. 95\% of the pixels in the radial range of $R = 1$-$7$ kpc are used.
The top panels of Figure \ref{fig:SNplot} 
show the S/N maps for $I_{\rm CO}$ and $\Sigma_{\rm HI}$.
The pixels with S/N$\le 3$ are blanked (white in these maps), and all of them are in the outermost annulus.
The bottom panel shows a correlation between the $I_{\rm CO}$ and $\Sigma_{\rm HI}$ S/N ratios. 
The S/N ratio of $\Sigma_{\rm HI}$ levels around $\sim$20-30, while that in $I_{\rm CO}$ reaches higher values, because the observed $\Sigma_{\rm HI}$ saturates around $10\,\Msun\,\pc^{-2}$, above which the gas phase is generally molecular \citep{Wong:2002lr, Bigiel:2008aa}.
In the lower S/N range ($\lesssim 20$), the S/N ratios in $I_{\rm CO}$ and $\Sigma_{\rm HI}$ are more or less comparable (i.e., one does not significantly dominate the other in a systematic way).
\citet{Sandstrom:2013vn} thoroughly discussed the capabilities and limitations of these fitting methods, including their dependence on the S/N 
of the tracers. The uncertainties here are small enough that the fitting methods, e.g. the $\Delta$log$\GDR$ scatter minimization approach, should give unbiased results.

\subsubsection{The $\chi^2$ Minimization} \label{sec:minch2}

Using eqs. (\ref{eq:fiteq0}) and (\ref{eq:Ico}),
a solution of ($\alphaCO$, $\GDR$) is derived
with the observed variables of ($\Sigma_{\rm HI}$, $I_{\rm CO}$, $\Sigma_{\rm d}$).
We rewrite eq. (\ref{eq:fiteq0}) as,
\begin{equation}
    \Sigma_{\rm HI}
        = \GDR \Sigma_{\rm d} - \alphaCO I_{\rm CO}.  \label{eq:fiteq3}
\end{equation}
and define $\chi^2$ as
\begin{equation}
    \chi^2 = \sum_{i}^{} \frac{(\Sigma_\mathrm{HI}-\GDR \Sigma_{\rm d}+\alphaCO I_{\rm CO})^2}
    {\sigma_{\Sigma_{\rm HI}}^2 + (\GDR \sigma_{\Sigma_{\rm d}})^2 + (\alphaCO\sigma_{I_{\rm CO}})^2}.
    \label{eqn:chi2_sd}
\end{equation}
$\sigma_{\Sigma_{\rm HI}}$, $\sigma_{I_{\rm CO}}$, and $\sigma_{\Sigma_{\rm d}}$ are
the errors in $\Sigma_{\rm HI}$, $I_{\rm CO}$, and $\Sigma_{\rm d}$ respectively.
The summations are over all image pixels, $i$, within a region of interest in the galaxy.

\begin{figure}
    \centering
    \includegraphics[width=0.45\textwidth]{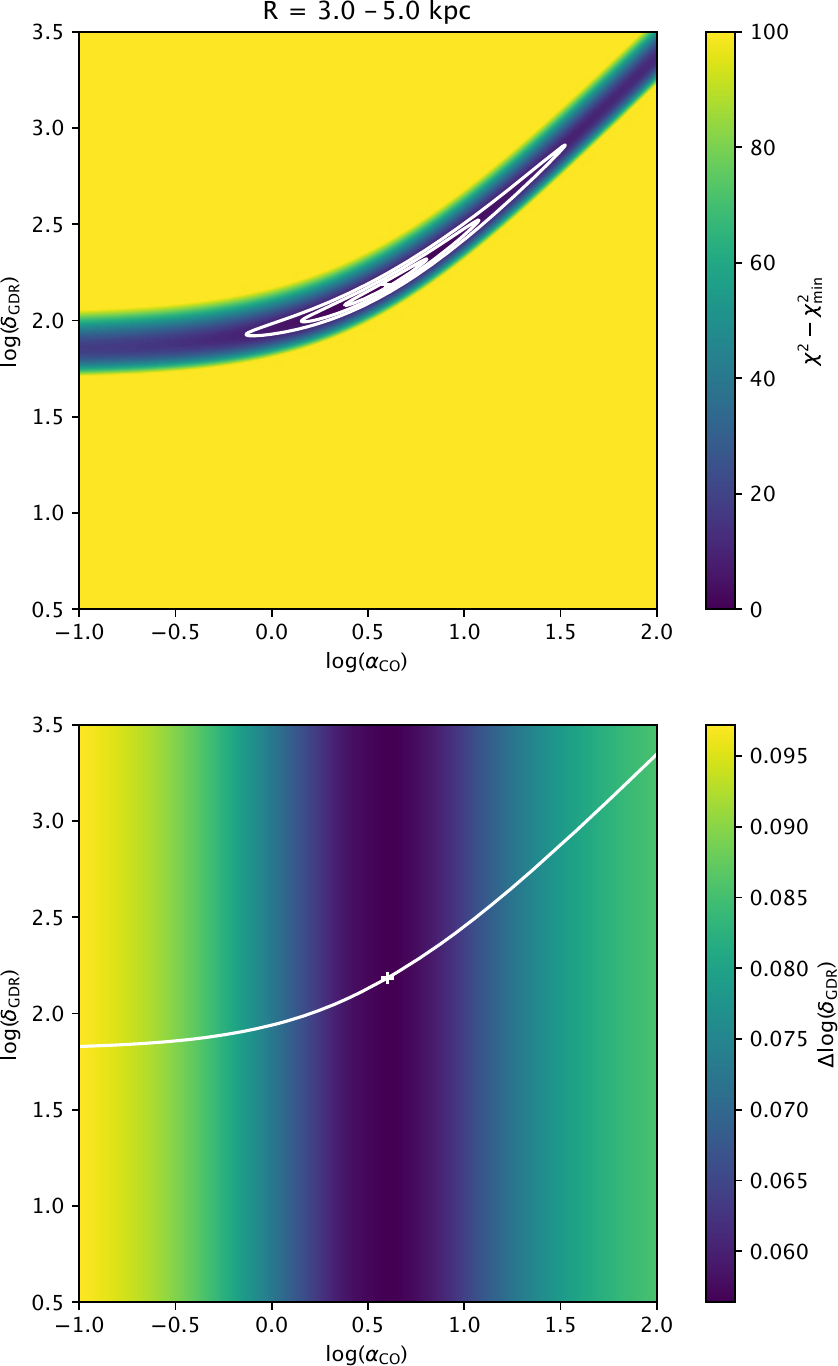}
    \caption{Example to show how solutions for $\alphaCO$ and $\delta_{GDR}$ are found simultaneously in the annulus from $R$ = 3-5 kpc, using the two methods.  
    Panel (a) shows $\Delta \chi^2 = \chi^2-\chi_{\rm min}^2$ for the $\chi^2$ minimization method.
    The contours are at $\Delta \chi^2=1.0$, 4.0, and 9.0,
    which correspond to the confidence levels of $1$, $2$, and $3\sigma$ when they are projected
    to the $\alphaCO$ or $\GDR$ axis, if the noises in the images and analysis follow a Gaussian distribution.
    Panel (b) shows the scatter in $\log \GDR$ for the $\Delta \log \GDR$ minimization method. 
    {The white solid line traces the average value of $\log \GDR$ derived
    when the scatter $\Delta \log \GDR$ is minimized at a fixed $\alphaCO$.}
    }
    \label{fig:acogdrspace}
\end{figure}

The $\chi^2$ values are calculated on a 2-dimensional grid of
($\alphaCO$, $\GDR$).
The minimum $\chi^2$ is searched in this space.
Figure \ref{fig:acogdrspace}a shows an example for an annulus of $R=3$-$5\kpc$.
If the errors follow the normal distribution,
the $\chi^2$ values can be converted to confidence intervals to assess the uncertainties (errors).
The figure shows three contours at $\Delta \chi^2 \equiv \chi^2-\chi_{\rm min}^2=1$, 4, and 9, which correspond to
the confidence levels (CLs) of 68, 95, and 99.7\% when they are projected to the $\alphaCO$ or $\GDR$ axis.
The results of the fitting are discussed in Section \ref{sec:resultsradial}.

\subsubsection{The $\log \GDR$ Scatter Minimization} \label{sec:mingdr}

A solution can also be derived by minimizing the scatter, $\Delta \log \GDR$,
in $\log \GDR$ over a region of interest.
For a fixed $\alphaCO$, we calculate a map of
\begin{equation}
    \GDR = \frac{\Sigma_{\rm g}}{\Sigma_{\rm d}} = \frac{\Sigma_{\rm HI}+\alphaCO I_{\rm CO}}{\Sigma_{\rm d}} \label{eq:gdrmap}
\end{equation}
and evaluate the scatter $\Delta \log \GDR$.
This is repeated for a range of $\log \alphaCO \in (-1.0, +2.0)$,
hence for a wide range of $\alphaCO \in (0.1, 100.0)$,
to find the $\alphaCO$ that minimizes the scatter.
Following \citet{Sandstrom:2013vn}, the scatter and mean of $\log \GDR$ are calculated
as the scale and location of the robust statistics, respectively.
\citet{Sandstrom:2013vn} showed, in their Figure 1c, that the minimization of $\Delta \log \GDR$
is equivalent to adopting a constant $\GDR$ (i.e., the inverse of their dust-to-gas ratio - DGR)
independent of gas phase (i.e., $I_{\rm CO}/\Sigma_{\rm HI}$, the ratio between $I_{\rm CO}$ and $\Sigma_{\rm HI}$).

Figure \ref{fig:acogdrspace}b shows an example for the same region ($R=3$-$5\kpc$) as in Figure \ref{fig:acogdrspace}a.
The color in the background shows $\Delta \log \GDR$ as a function of $\alphaCO$: each $\alphaCO$ corresponds to one value of $\Delta \log \GDR$, and hence,
$\Delta \log \GDR$ changes only in the horizontal direction.
In addition, an average $\log \GDR$ at each $\alphaCO$ can be evaluated with eq. (\ref{eq:gdrmap}).
We adopt the biweight location to approximate the average
following \citet{Sandstrom:2013vn}, which is shown as the white line in Figure \ref{fig:acogdrspace}b.
Along this line, the point at the minimum of $\Delta \log \GDR$, in the color background, is the solution.
The result of the fitting is discussed in Section \ref{sec:resultsradial}.

\section{Results} \label{sec:results}

\subsection{Radial Gradients} \label{sec:resultsradial}

We search for solutions in each annulus with a width of 2~kpc and increment of 1~kpc. 
The width is selected so that it is greater than the $56.6\arcsec$ resolution, which is 1.23~kpc at the distance of M83.
Adjacent annuli overlap and are not independent from each other, but are useful in finding a radial trend.
The central 1~kpc in radius is excluded as $\Sigma_{\rm HI}$ cannot be measured accurately, due to the bright continuum emission at the galactic center, and HI absorption.

Table \ref{tab:newresults} summarizes the derived ($\alphaCO$, $\GDR$) solutions for each fitting method and annulus.
Overall, the ($\alphaCO$, $\GDR$) values obtained between the fitting methods for the same annulus are similar.
The values within the parentheses show the minimum and maximum $\alphaCO$ and $\GDR$ of the $\Delta \chi^2=1$ region, which would correspond to the 68\% CL (i.e., $1\sigma$) if all 
errors in this analysis follow the Gaussian statistics entirely, and gives some gauge of the error in this analysis.
Under this treatment, at least for the $\chi^2$ method, the overall errors in $\alphaCO$ in each annulus is about a factor of 2. 

\begin{figure}[h]
    \centering
    \includegraphics[width=0.4\textwidth]{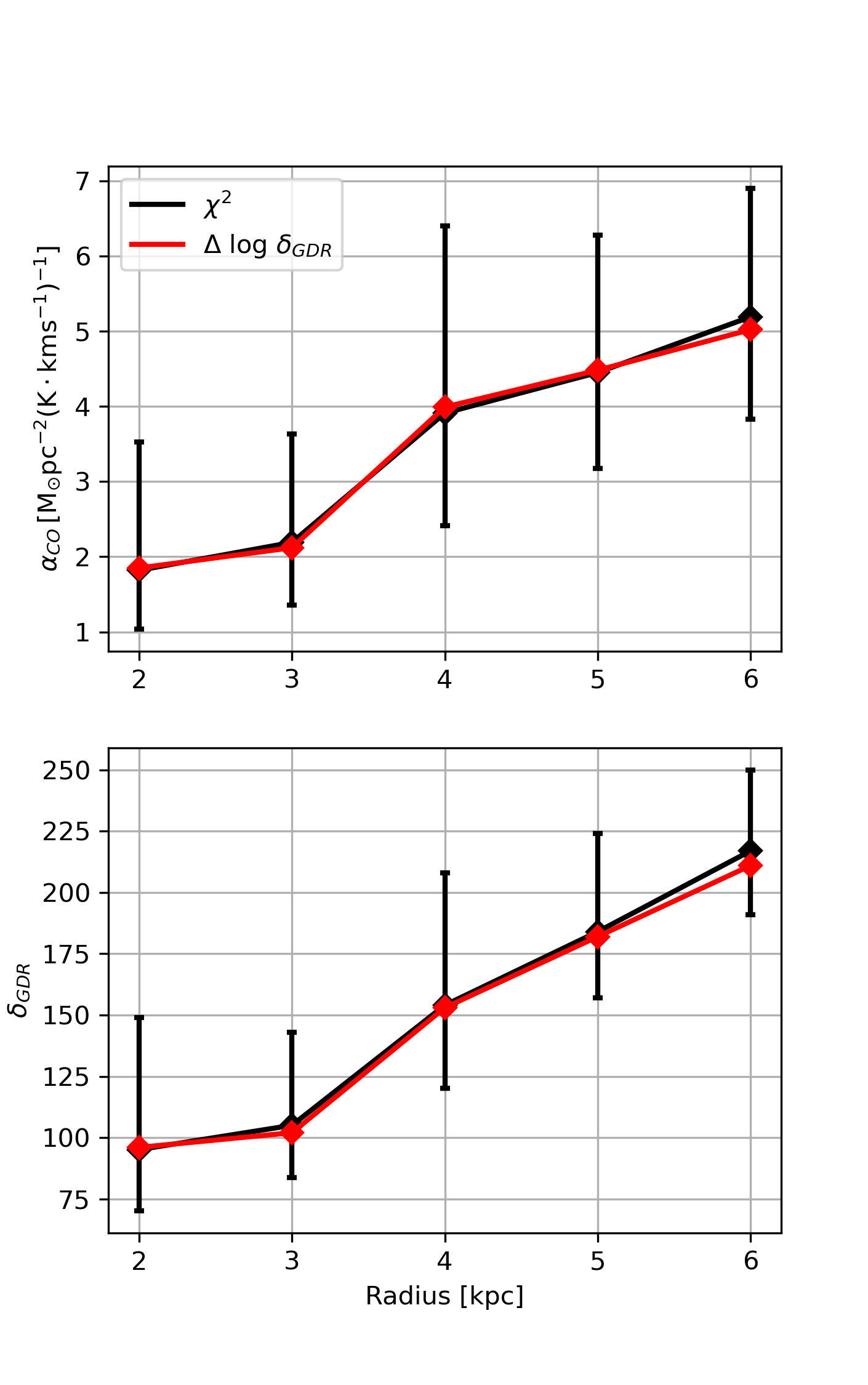}
    \caption{Radial variations of $\alpha_{CO}$ and $\GDR$ (Table \ref{tab:newresults}) plotted at the midpoint of each annulus.}
    \label{fig:radial_var}
\end{figure}

Figure \ref{fig:radial_var} shows the radial variations of $\alphaCO$ and $\GDR$.
All the solutions suggest that both $\alphaCO$ and $\GDR$ increase as a function of radius, independent of the fitting method. 
If we take the results from the $\chi^2$ method (column 4), 
$\alphaCO$ monotonically increases, by a factor of 2-3, from $\alphaCO\sim$1.82 (1.04-3.53 at 68\% CL) in $R=$ 1-3~kpc to $\sim$5.19 (3.83-6.90) in $R=$ 5-7~kpc.
The units of $\alphaCO$, $\Msun\,\pc^{-2}[{\rm K} \cdot \kmps]^{-1}$, are omitted for simplicity.

$\GDR$ also increases by a factor of 2 from $\sim 95$ (70.2, 149) in $R=$ 1-3~kpc to $\sim217$ (191, 250) in $R=$ 5-7~kpc (column 5).
Again, both methods show the same radial trend in $\GDR$. 
Eq. (\ref{eq:fiteq0}) indicates that the absolute value of $\GDR$ is sensitive to the normalization of the scale of $\Sigma_{\rm d}$, which depends on the dust model. For example, if $\Sigma_{\rm d}$ is overestimated by, say, a factor of 2, $\GDR$ would be underestimated by the same factor.
Hence, the absolute value of $\GDR$ carries such systematic uncertainties.

Here, we assumed that $\alphaCO$ and $\GDR$ are constant in each annulus.
There is discussion on the metallicity-dependence of $\alphaCO$ and $\GDR$ \citep{Arimoto:1996aa, Israel:1997lr, Leroy:2011lr, Schruba:2012aa, Remy-Ruyer:2014aa, Hunt:2015aa}.
The metallicity in M83 has been measured in several studies \citep[e.g., ][]
{Bresolin:2009ce, Pilyugin:2014aa, Grasha:2022aa}, and their absolute values are not necessarily consistent with each other, often due to different calibrations.
However, the azimuthal variation, which is seen as the scatter at each radius, is consistently small in all the studies.
Therefore, it seems safe to assume that the metallicity is approximately constant in each annulus,
and thus that $\alphaCO$ and $\GDR$ do not suffer much from metallicity-dependent variations within each annulus. We find a change of $\Delta$(12 + log(O/H)) = -0.06 within one annulus using the radial gradients discussed in Section \ref{sec:metaldependence}.

\begin{deluxetable*}{ccccccccc}
\tablecaption{Results using $\Sigma_{\rm HI}$, $I_{\rm CO}$, and $\Sigma_{\rm d}$ at $56.6\arcsec$ resolution
with the $\chi^2$ and $\Delta \log \GDR$ minimization methods.\label{tab:newresults}}
\tablehead{
\colhead{(1)} & 
\colhead{(2)} & \colhead{(3)} & \colhead{(4)} & \colhead{(5)} & \colhead{} & \colhead{(6)} &
\colhead{(7)} \\
 \cline{1-9} \colhead{} & 
\colhead{} & \colhead{} &  \multicolumn{2}{c}{$\chi^2$}  & \colhead{}  & \multicolumn{2}{c}{$\Delta \log \GDR$} & \colhead{} \\  
\cline{4-5} \cline{7-9} 
\colhead{Radius} & 
\colhead{$N_{\rm beam}$} &  \colhead{$L^{\rm tot}_{\rm CO}$} & \colhead{$\alphaCO$} & \colhead{$\GDR$} & \colhead{} & \colhead{$\alphaCO$} & \colhead{$\GDR$} & \colhead{} } 
\startdata
  1.0-3.0  & 13.0 & 3.1e8 & 1.82 (1.04, 3.53)  & 95.1 (70.2, 149) &&  1.85  & 96.0   \\ 
  2.0-4.0  & 19.5 & 2.9e8 &  2.19 (1.36, 3.63)   &  105 (83.6, 143)  && 2.12 &  102   \\   
  3.0-5.0  & 26.3 & 2.3e8 & 3.91 (2.41, 6.40) &  154 (120, 208) && 3.99 & 153  \\    
   4.0-6.0  & 32.7 & 1.9e8 &  4.45 (3.17, 6.28) & 184 (157, 224) && 4.48 & 182  \\
 5.0-7.0  & 35.1 & 1.2e8 &  5.19 (3.83, 6.90)   & 217 (191, 250)   &&  5.02 &  211      \\
  \cline{1-9}
   Luminosity-weighted & -- & -- & 3.14 (2.06, 4.96) & 137 (111, 182) && 3.13 & 136  \\
  Area-weighted  & -- & -- & 3.96 (2.73, 5.81) & 167 (140, 209)  && 3.93 & 164 \\ 
  \cline{1-9} 1.0-7.0  & 74.4 & 6.7e8 & 6.03 (5.19, 6.90)  &  221 (203, 241)   &&   5.95  & 214    
\enddata
\tablecomments{
(1) Radial range of aperture for fit in kpc.
(2) Number of independent beams in aperture.
(3) Total CO luminosity in corresponding annulus with units {$[{\rm K} \cdot \kmps] \cdot \pc^{2}$}. 
(4)(6) CO-to-H$_2$ conversion factor including the contribution of helium (a factor of 1.36) in {$\Msun\,\pc^{-2}[{\rm K} \cdot \kmps]^{-1}$}.
(4)(5) The two values within the parenthesis are the minimum and maximum values within $\chi^2 = \chi_{\rm min}^2+1$.
(6)(7) Gas-to-dust mass ratios. }
\end{deluxetable*}

\subsection{Galaxy-wide Averages} \label{sec:resultsaverage}

The average $\alphaCO$ and $\GDR$ give the representative values of the whole galaxy.
Since both of these parameters vary systematically along the galactic radius,
their averages depend on how the annuli are weighted in averaging. 
Here we calculate the galaxy-average values by applying
different weightings on the annulus-based solutions (Table \ref{tab:newresults})
or by applying the fitting methods in Section \ref{sec:findsolutions} to the whole galaxy.
The best averaging scheme may depend on the purpose of usage.
For example, to derive the total molecular gas mass over a galaxy, the luminosity-weighted average $\alphaCO$ may be the best.

We first calculate the averages from the annulus-based analysis (Table {\ref{tab:newresults})}. 
With the $\chi^2$ method, the averages weighting by luminosity and area are 3.14 (2.06 - 4.96) and 3.96 (2.73 - 5.81), respectively, with the values within the parenthesis indicating the 68\% confidence points.
With the $\Delta {\rm log} \GDR$ method, the average $\alphaCO$ are 3.13 and 3.93 for the luminosity and area weightings, respectively. 
Since the luminosity is concentrated more towards the inner galaxy,
the luminosity-weighted galaxy average tends to weight the inner part of the disk more, and results in a smaller $\alphaCO$.
On the other hand, there is more area in the outer galaxy, and the area-weighted average skews more towards the values of the outer disk,
leading to the larger $\alphaCO$.
Nevertheless, all these averages are consistent within an expected factor of 2 uncertainty.

The luminosity and area weighted $\GDR$ using the $\chi^2$ method are 137 (111, 182) and 167 (140, 209), while the $\Delta$log$\GDR$ method gives 136 and 164. 
The $\GDR$ increases with radius, and the luminosity weighting weighs the inner disk more, while the area weighting weighs the outer disk more.

We also apply the $\chi^2$ and $\Delta \log \GDR$ methods to the whole area ($R=$ 1-7$\kpc$; excluding the $R<1\kpc$ area).
The results are in the last row of {Table \ref{tab:newresults}}.
The $\chi^2$ fitting gives $\alphaCO=6.03$ with $\Delta \chi=1$ range of 5.19-6.90.
The minimization of $\log \GDR$ also gives a very similar value
of $\alphaCO=5.95$. 
These values are almost a factor of 2 larger than the annulus-based averages,
and are likely due to the non-linearity between $I_{\rm CO}$ and the molecular gas surface density 
$\Sigma_{{\rm H}_2}$, calculated as the difference between
the dust-based total gas and atomic gas surface densities (i.e., $\Sigma_{\rm H_2}=\GDR \Sigma_{\rm d} - \Sigma_{\rm HI}$; see Section \ref{sec:radgrad}). In other words, the assumed linear fit equation (\ref{eq:fiteq3}) may break down on the scales of the entire galactic disk in M83, as most environmental parameters (such as metallicity, cloud populations, dominant gas phase - molecular vs. atomic) can vary across the disk.

This is a potential limitation of this method, but in our annulus-based analysis in Section \ref{sec:resultsradial}, this limitation is mitigated.

\citet{Bolatto:2013ys} summarized the various measurements of $\alphaCO$ in the MW and recommended $\alpha_{\rm MW}=\alphaCO=4.35 \Msun\,\pc^{-2}[{\rm K}  \kmps]^{-1}$ with uncertainties of $\pm$30\% 
within the inner MW disk (R=1-9 kpc) and a factor of 2 for ``normal" galactic disks with solar metallicity.
Taking into account our own uncertainties,
the derived $\alphaCO$ averages are consistent within the $\alphaCO$ in the solar metallicity ($\alphaCO \sim$ 2.2-8.7, i.e., $1/2\alpha_{\rm MW}$ to $2\alpha_{\rm MW}$).

\section{Discussion}

\subsection{Previous Measurements in M83}
\label{sec:comparisons}

In this study, we derived $\alphaCO$ and $\GDR$, simultaneously in M83.
Previous studies of this galaxy derived $\alphaCO$ by fixing $\GDR$ \citep{Jiao:2021aa}, derived $\GDR$ by fixing $\alphaCO$ \citep{Foyle:2012aa}, or derived $\alphaCO$ and $\GDR$ together \citep{Yasuda:2023aa}.

\citet{Jiao:2021aa} estimated the CO-to-H$_2$ and CI-to-H$_2$ conversion factors in M83, also using the dust-based method as we do in this study (eq. \ref{eq:fiteq0}), but with THINGS HI and Nobeyama CO(1-0) data \citep{Walter:2008mw, Kuno:2007zr} and the \citet{Draine:2007fj} dust model.
They fixed $\GDR$, assuming three forms of metallicity-dependent $\GDR$, and derived
$\alpha_{CO}$ finding an increase from $\sim$1.0 to $\sim$3.2 from the central region to their CO detection radius limit of $R \sim 4$ kpc (their Figure 2).
These appear consistent with the results derived in Section \ref{sec:results}, but with caution. As demonstrated in Figure \ref{fig:archivedata}, the ratios of the archival HI and CO(1-0) fluxes used in their study over our fluxes are, on average, only 40 and 80\%, respectively.
These ratios also vary radially and azimuthally, e.g. between the arm and interarm regions.
If we simplistically take the average differences and compare them with eq. (\ref{eq:fiteq0}),
their $\ICO/\Sigma_{\rm HI}$ ratio would be a factor of two larger than ours, which should result in twice smaller $\alphaCO$ than ours.
The apparent consistency may indicate that the dust-based method, which both their and our studies employed, has an intrinsic uncertainty of a factor of two. 

\citet{Yasuda:2023aa} analyzed the same HI and CO(1-0) data as \citet{Jiao:2021aa} and adopted the SED fitting by \citet{Casey:2012ab} for $\Sigma_{\rm d}$.
They derived both $\alphaCO$ and $\GDR$ together.
They obtained an average of $\alphaCO=0.84$ for $R\lesssim 4.7\kpc$ (0.60 - 1.23, within the 68\% confidence level in the $\alphaCO$-$\GDR$ plane),
and a radial increase in $\alphaCO$ by about a factor of two from 0.34 (0.18 - 0.80) to 0.78 (0.43 - 1.42) between the inner ($R<1.7\kpc$) and outer ($>1.7\kpc$, up to $R\sim4.7\kpc$) disk.
The radial increase is consistent with the factor of two increase in our results, from 1.82 to 3.91 (Table \ref{tab:newresults}).
The values of their $\alphaCO$ are about 
twice smaller than our results, which can be explained by the flux errors in their HI and CO data (as discussed above).
\citet{Yasuda:2023aa} also found $\GDR=$73 (59-94) on average, with an increasing trend from 33 (21-66) in $R<1.7\kpc$ to 72 (55-102) in $>1.7\kpc$.
These are about three times smaller than our results (Table \ref{tab:newresults}).
This discrepancy could also be explained by the factor of 2.5 smaller HI flux in their data,
which reduces $\GDR$ by the same factor (see eq. \ref{eq:fiteq0}).

\citet{Foyle:2012aa} fixed $\alphaCO$ and derived $\GDR$ on spatial scales of $\sim 0.8 \rm kpc$. They used a modified blackbody fit for $\Sigma_{\rm d}$ and applied both a constant (MW value) and metallicity-dependent $\alphaCO$. 
They used THINGS HI and CO(3-2) emission from the JCMT for the atomic and molecular gas, respectively.
From their Figure 10, the radial $\GDR$ profile looks roughly constant, with $\delta_{GDR}$ values of $\sim$ 80 and $\sim$ 60 when using a constant and metallicity-dependent $\alphaCO$ (their equation 6) respectively. These values could be corrected upwards, possibly by a factor of two, if the factor of 2 flux error in the THINGS HI data is taken into account (see eq. \ref{eq:fiteq3}). To that end, the $\delta_{GDR}$ derived in this study are consistent with their values.

\subsection{Possible Origin of Radial Gradients}\label{sec:radgrad}
We find that $\alphaCO$ and $\GDR$ systematically increase with galactic radius (Section \ref{sec:resultsradial}).
Such variations are often attributed to the dependence on metallicity \citep{Wilson:1995lr, Arimoto:1996aa, Israel:1997lr, Leroy:2011lr}, and we discuss their possible metallicity-dependence in Section \ref{sec:metaldependence}.
At the same time, many galactic parameters vary radially (e.g., stellar density and radiation field, gravitational potential, strength of galactic structures, cosmic ray intensity, gas amount, and phase, as well as metallicity),
and the $\alphaCO$ and $\GDR$ could correlate with any of those radially-variable parameters. 
It is 
valuable to discuss other possible origins of the $\alphaCO$ and $\GDR$ variations.
In Section \ref{sec:clouddependence}, we discuss another possibility: a dependence on the underlying molecular cloud populations.
Identifying the exact origin is beyond the scope of this work, and here we aim to investigate a possible origin.

\begin{figure}[h]
    \centering
     \includegraphics[width=0.45\textwidth]
    {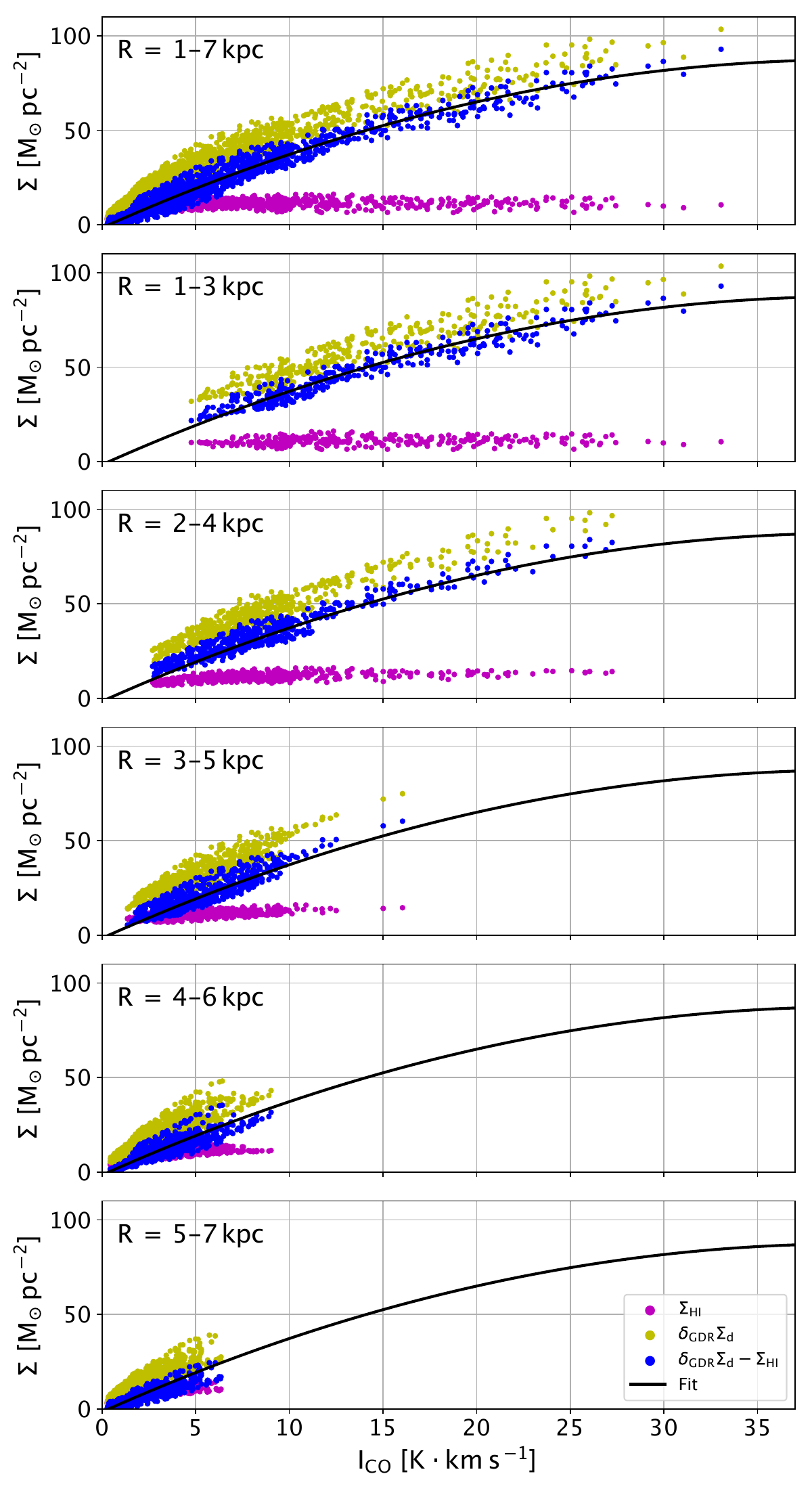}
    \caption{Plots of three surface densities as a function of $I_{\rm CO}$.
    The three include the HI gas surface density, ``$\Sigma_{\rm HI}$" (magenta),
    the total gas surface density traced by dust, ``$\GDR\Sigma_{\rm d}$" (yellow), and
    the H$_2$ surface density calculated as a difference of the dust-based total gas and HI, ``$\GDR\Sigma_{\rm d}-\Sigma_{\rm HI}$" (blue).
    All the surface densities include the contribution of He mass (a factor of 1.36).
    The panels correspond to, from top to bottom, the radial ranges of $R=$1-7 (whole area), 1-3, 2-4, 3-5, 4-6, and 5-7~kpc.
    For this example, we set $\GDR=137$, the luminosity weighted average for $\GDR$. The plots for $\GDR$=167 and 221, which correspond to the values for the area weighted average and over the entire disk ($R=$ 1-7~kpc), show similar curved shapes.
    The black solid line is a 2-d polynomial fit to the blue points in the top panel,
    and between the $x$ and $y$ axes, it is 
    $y= -0.055 x^2 + 4.43 x - 1.54$. 
    Note that the slope of ``$\GDR\Sigma_{\rm d}-\Sigma_{\rm HI}$" (blue points) vs. $I_{\rm CO}$ is $\alphaCO$.}
    \label{fig:SurfDIco}
\end{figure}

Figure \ref{fig:SurfDIco} presents a clue to understanding the radial gradient of $\alphaCO$.
This plot shows molecular and atomic gas surface densities vs. $I_{\rm CO}$. The top panel is for the disk ($R=$1-7$\kpc$), and the subsequent panels show the same data, but for each $2\kpc$-width annulus 
separately. 
If $\alphaCO$ is constant, then we expect a straight line for the molecular gas, with $\alphaCO$ as the slope. 
The blue points ``$\GDR\Sigma_{\rm d}-\Sigma_{\rm HI}$" represent the molecular gas surface density
(equivalent to $\Sigma_{\rm H_2}$ when $\GDR$ is calibrated; see eq. \ref{eq:fiteq0}).
Again, $\alphaCO=\Sigma_{\rm H_2}/I_{\rm CO}$ is the ratio between the $y$-axis and the $x$-axis in this plot
and is the CO-to-H$_2$ conversion factor,
when the $y$-intercept is zero. 
For comparison, the figure also includes the total HI+H$_2$ gas surface density ``$\GDR\Sigma_{\rm d}$" (yellow)
and $\Sigma_{\rm HI}$ (magenta).

The blue points show non-linear (curved) distributions.
The black solid lines are 2-d polynomial fits to the blue data points for $R=$1-7$\kpc$ (top panel) and are the same in all panels for reference. 
The curved distributions indicate that the slope $\alphaCO$ decreases with increasing $I_{\rm CO}$.
At the same time, these panels also show that the range of $I_{\rm CO}$ that each annulus occupies declines radially,
and hence, $\alphaCO$ increases with galactic radius.

This suggests that if the non-linear dependence of $\alphaCO$ on $I_{\rm CO}$ reflects intrinsic gas properties,
the range of $I_{\rm CO}$ that each radius covers determines the average $\alphaCO$ in that radius.
The non-linearity thus could explain the radial increase of $\alphaCO$.
\footnote{This radial trend (in Figure \ref{fig:SurfDIco}) would be amplified further if the radially-increasing $\GDR$ is also taken into account:
in a relative sense, the yellow data points ($\GDR\Sigma_{\rm d}$) would move downwards in smaller radii and upwards in larger radii, and hence,
the blue data points would also move downwards (in the lower $\alphaCO$
direction) in smaller radii and upwards (in the larger $\alphaCO$ direction) in larger radii.}

The above discussion relies on an assumption that the other parameters,
especially $\Sigma_{\rm d}$, are determined accurately.
It is also possible that the determination of $\Sigma_{\rm d}$ could suffer from some systematic errors
even after significant efforts, due to the inherent difficulty
in constraining dust properties.
The relation between $I_{\rm CO}$ and $\GDR\Sigma_{\rm d}$ (note that $\GDR$ is constant in the plots) is already non-linear in a wide range of $I_{\rm CO}$,
which is inherited by the molecular gas surface density ``$\GDR\Sigma_{\rm d}-\Sigma_{\rm HI}$" (blue).

\subsubsection{Metallicity Dependence} \label{sec:metaldependence}

Metallicity, as well as many other parameters, is likely to change radially,
and $\alphaCO$ and $\GDR$ could depend on any of those parameters \citep{Arimoto:1996aa, Israel:1997lr, Leroy:2011lr, Schruba:2012aa, Remy-Ruyer:2014aa, Hunt:2015aa, Accurso2017, Madden:2020aa, Chiang:2021aa} (see Appendix \ref{sec:metallicity}).
There are also theoretical explanations on the metallicity dependence of $\alphaCO$ \citep{Wolfire2010, Gong:2020aa, Hu2022}.
Recent works have emphasized the importance of metallicity for molecular gas mass estimations even within a single galaxy. \citet{Evans:2022aa} demonstrated that the $\alphaCO$ variations with metallicity have large effects on star formation efficiency estimates in the MW. 
Metallicity determinations with optical line emission
have systematic uncertainties. 
Previous studies have measured the metallicity in M83 with roughly consistent results \citep[e.g., ][]{Bresolin:2009ce, Pilyugin:2014aa, Grasha:2022aa}, but with some differences due to their different calibrations.
Hence, we focus primarily on the radial changes (differentials) in $\alphaCO$ due to radial metallicity changes, and secondarily on their absolute values.

We adopt the metallicity measurements of M83 by \citet{Bresolin:2009ce}.
They found a radial gradient of $12 + \rm{log}\left( {\rm O/H} \right) = 8.77(\pm 0.01) - 0.25(\pm 0.02) R/R_{25}$ (where $R_{25}=6.44\arcmin$) with the \citet{Bresolin:2007aa} calibration of metallicity measurements (hereafter, B07),
and the same radial gradient with a 0.47~dex higher oxygen-abundance with the \citet{Kewley:2002aa} calibration (KD02).
With this gradient,
the metallicity of M83 decreases radially by $\Delta ({\rm 12 + log(O/H)}) = -0.12$ from $R=$ 2 to 6~kpc (the midpoint of each annulus).
The corresponding increases in $\alphaCO$ are listed in Table \ref{tab:modelaco}
and are factors of $\sim1.2$ and $\sim1.3$, for the $\alphaCO$-metallicity relations of \citet[][W95]{Wilson:1995lr} and \citet[][A96]{Arimoto:1996aa} respectively,
when their calibrations are scaled for $\alpha_{{\rm MW}}$ = 4.35. For reference, the A96 and W95 metallicity relations have power-law slopes of -1.0 and -0.7 respectively.
The amount of radial change expected from the metallicity gradient can partially explain the radial gradient of $\alphaCO$ derived in this study.

The \emph{absolute} value of $\alphaCO$ suffers from uncertainties in the calibration of metallicity measurements.
From $R=$ 2 to 6~kpc,
the B07 metallicity calibration gives an increase of $\alphaCO$ = 5.66 to 6.80 for the W95 $\alphaCO$-metallicity relation, and $\alphaCO$ = 8.45 to 11.10 for the A96 relation.
The KD02 calibration gives $\alphaCO$ = 2.70 to 3.24 for the W95 relation,
and $\alphaCO$ = 2.80 to 3.68 for the A96 relation.
Table \ref{tab:modelaco} summarizes these $\alphaCO$ in columns 2-5.
The $\alphaCO$ values from the KD02 calibration are consistent with the results of this study, while those from the B07 calibration are about a factor of 2 larger
(see Table \ref{tab:newresults}).
Figure \ref{fig:overplotavgacov1} plots the metallicity-dependent $\alphaCO$ variations with respect to our results.
The bottom panel is the same as the top, but shows only the $\alphaCO$ from the KD02  calibration.

\subsubsection{Cloud Population Dependence} \label{sec:clouddependence}

As discussed in Section \ref{sec:radgrad} (see also Figure \ref{fig:SurfDIco}), $\alphaCO$ appears to depend on the range of $I_{\rm CO}$ that each annulus traces, and hence depends on the range of $\Sigma_{\rm H_{2}}$. It therefore seems natural to consider variations of the physical conditions of the gas, i.e., the average properties of the underlying molecular cloud populations, as another potential origin of the $\alphaCO$ variation.
\citet{Dickman:1986aa} calculated $\alphaCO$ as an average over a cloud population (see Appendix \ref{sec:Dickman}). We extend this idea and examine an effect of varying cloud populations on $\alphaCO$. 
While our $56.6\arcsec$ (1.23~kpc) resolution does not resolve individual molecular clouds,
it has been known that the cloud population in M83 varies radially: massive (luminous) clouds are populated more in the inner disk than in the outer disk \citep[][Hirota et al. in prep.]{Freeman:2017aa}.

The dynamical state of molecular clouds is often characterized with the virial parameter $\alpha_{\rm vir}$, 
which is the ratio between the kinetic energy $K$ and gravitational potential energy $U$ of cloud \citep{Bertoldi:1992lr},
\begin{equation}
    \alpha_{\rm vir} \equiv \frac{2K}{|U|}. \label{eq:virdef0}
\end{equation}
A cloud is gravitationally-bound when $\alpha_{\rm vir}\leq 2$ (i.e., $K \leq |U|$), and is in virial equilibrium when $\alpha_{\rm vir} = 1$ ($2 K = |U|$, the virial theorem).
Hence, gravitationally-bound clouds should show $\alpha_{\rm vir}= $1-2.

For a spherical cloud with the mass $M$, line-of-sight velocity dispersion $\sigma_{\rm v}$, and density profile $\rho(r)\propto r^{-\xi}$ truncated at $r=R$, we have $U=-(3-\xi)/(5-2\xi) (G M^2/R)$ and $K=3/2 (M \sigma_{\rm v}^2)$.
Thus, with a definition of the virial mass $M_{\rm vir} = 3(5-2\xi)/(3-\xi) (R \sigma_{\rm v}^2/G)$,
\begin{equation}
    \alpha_{\rm vir} = \frac{M_{\rm vir}}{M}. \label{eq:virdef}
\end{equation}
Molecular clouds in the Milky Way (MW) show central concentrations ($\xi>0$) but do not have as steep a profile as the isothermal sphere ($\xi<2$). If we adopt $\xi=1$ as done by \citet{Solomon:1987pr}, \textbf{$M_{\rm vir} = 9/2 (R \sigma_{\rm v}^2/G)$.\footnote{We take this definition of $M_{\rm vir}$ so that the virialized state is represented by $\alpha_{\rm vir}=1$, since this seems to be a natural definition of the virial parameter. Some studies define $M_{\rm vir}^{\prime} \equiv 5 R \sigma_{\rm v}^2/G$ 
independent of $\xi$ and 
in turn their virial parameter is $\alpha_{\rm vir}^{\prime} \equiv M_{\rm vir}^{\prime}/M = (5/3)((3-\xi)/(5-2\xi))\alpha_{\rm vir}$. In their case, the virialized state is $\alpha_{\rm vir}^{\prime}$=1.00, 1.11, and 1.67 for $\xi$= 0, 1, and 2, respectively.}}

The true cloud mass $M$ is often represented by the luminous mass $M_{\rm lum}$($=\alphaCO L_{\rm CO} $) using the cloud CO luminosity $L_{\rm CO}$, and hence,
\begin{equation}
    \alpha_{\rm vir} \alphaCO = \frac{M_{\rm vir}}{L_{\rm CO}}. \label{eq:viraco1}
\end{equation}

\citet{Solomon:1987pr} analyzed a population of molecular clouds with masses $>10^4\Msun$ in the MW 
and found a power-law relation of $M_{\rm vir}=39 L_{\rm CO}^{0.81}$ with $M_{\rm vir}$ and $L_{\rm CO}$
in units of $\Msun$ and $\rm K\,\kmps\,\pc^2$, respectively.
Using the same data but conducting a separate analysis,
\citet{Scoville:1987lp} obtained a linear relation, $M_{\rm vir}=7.9L_{\rm CO}$, for clouds with $>10^5\Msun$.
In a generalized form, this empirical relation, along with eq. (\ref{eq:viraco1}), is
\begin{equation}
    \alpha_{\rm vir} \alphaCO = \frac{M_{\rm vir}}{L_{\rm CO}} = \alpha_{\rm ref} \left( \frac{L_{\rm CO}}{L_{\rm ref}} \right)^{\beta}.  \label{eq:viraco2}
 \end{equation} 
We adopt a CO reference luminosity of $L_{\rm ref}=10^5\,\rm K\cdot \kmps \pc^2$. 
The relations found by \citet{Solomon:1987pr} and \citet{Scoville:1987lp} translate to $\beta = -0.19$ and $\beta=0$, respectively.

In the case of $\beta \neq 0$, eq. (\ref{eq:viraco2}) means that $\alphaCO$ and/or $\alpha_{\rm vir}$ change with the cloud luminosity.
It has been discussed in two ways, by assuming either $\alphaCO$ or $\alpha_{\rm vir}$ to be constant.
\citet{Chevance:2022aa} assumed a constant $\alphaCO$ and discussed that $\alpha_{\rm vir}$, the dynamical state of clouds, changes with cloud luminosity, and hence with mass.
\citet{Bolatto:2013ys} adopted $\alpha_{\rm vir}=1$ (all clouds are in virial equilibrium) and explained that
$\alphaCO$ decreases with cloud luminosity and mass.
If $\alpha_{\rm vir}$ is constant, the average $\alphaCO$ in a region should depend on the underlying cloud population.

In order to evaluate the average $\alphaCO$ in each annulus,
we adopt the simplest form of commonly used luminosity functions, a truncated power-law cloud luminosity function \citep{Sanders:1985ud, Solomon:1987pr, Rosolowsky2005,Roman-Duval:2010fk, Colombo:2014aa}
defined between a maximum and minimum cloud luminosity, $L_{\rm max}$ and  $L_{\rm min}$
\begin{eqnarray}
    \Phi(\LCO){\rm{d}}\LCO = \Phi_{\rm max}\left(\frac{\LCO}{L_{\rm max}} \right)^\gamma {\rm{d}}\LCO, \label{eq:lumfunc}
\end{eqnarray}
where $\Phi(\LCO)\,{\rm{d}}\LCO$ is the number of clouds with luminosity $\LCO$ in a volume and
$\Phi_{\rm max}$ is the number density of clouds with $L_{\rm max}$. 
We adopt a luminosity function instead of a mass function because a mass function would require assuming a fixed conversion factor.

The total luminosity and mass in an annulus are
$L_{tot}=\int \LCO \Phi \, {\rm d}\LCO/\int \Phi \, {\rm d}\LCO$ and
$M_{tot}=\int \alpha_{CO} \LCO \Phi \, {\rm d}\LCO/\int \Phi \, {\rm d}\LCO$, respectively.
With eq. (\ref{eq:viraco2}) and constant $\alpha_{\rm vir}$, we derive the average $\langle \alphaCO \rangle= M_{tot}/L_{tot}$ for the cloud population,
\begin{eqnarray}
    &\langle \alphaCO \rangle &= \left( \frac{1}{\alpha_{\rm vir}} \right) \alpha_{\rm ref} \frac{\gamma+2}{\beta+\gamma+2} \left( \frac{L_{\rm max}}{L_{\rm ref}}\right)^{\beta}     \label{eqn:acoPL}\\
     &\times& \left[ \left( 1 - \left( \frac{L_{\rm min}}{L_{\rm max}} \right)^{\beta + \gamma + 2} \right) / \left( 1 - \left(\frac{L_{\rm min}}{L_{\rm max}}\right)^{\gamma + 2} \right) \right].  \nonumber 
\end{eqnarray}
Note that $\alpha_{\rm ref}$ is in the units of $\alphaCO$, and $\alpha_{\rm vir}$ is a dimensionless parameter.
$L_{\rm min}/L_{\rm max}$ is small \citep[$\sim 10^{-2}$, ][]{Solomon:1987pr, Scoville:1987lp}, and 
the last term within the bracket is around 1.
Thus, in the limit where $L_{\rm min}/L_{\rm max} \rightarrow 0$,
$\langle \alphaCO \rangle$ depends on the luminosity of the brightest and hence most massive cloud.

Hirota et al. (in prep.) generated a catalog of molecular clouds in M83. In their analysis, $\sim$91\% of the total CO luminosity within the mapped area was assigned to molecular clouds.
They fitted a truncated power-law cloud luminosity function and obtained ($\gamma$, $L_{\rm max}$) in each annulus (Table \ref{tab:clumfunc}).
Using their cloud catalog, we 
determine ($\alpha_{\rm ref}$, $\beta$) of eq. (\ref{eq:viraco2})
in Appendix \ref{sec:m83fits}.
These parameters are listed in Table \ref{tab:arefmodels} in the appendix.
In this catalog, small clouds are not entirely spatially resolved.
Hirota et al. (in prep.)
uses their peak brightness temperature $T_{\rm peak}$ as a rough measure to separate likely-resolved clouds  ($T_{\rm peak}>2$~K) from likely-unresolved ones  ($<2$~K) (Appendix \ref{sec:m83fits}).
Therefore, we obtained ($\alpha_{\rm ref}$, $\beta$)
for all clouds (denoted ``M83$_{\rm all}$") and likely-resolved clouds with $T_{\rm peak}>2$~K (``M83$_{\rm >2K}$").
As an example of unambiguously-resolved clouds, we also use the clouds in the MW \citep[``MW"; ][]{Solomon:1987pr}.
For consistency, we made our own fit to the MW clouds (Appendix \ref{sec:m83fits})
and use the MW relation for the M83 clouds as well.

\begin{figure*}[t]
    \centering
    \includegraphics[width=1.0\textwidth]{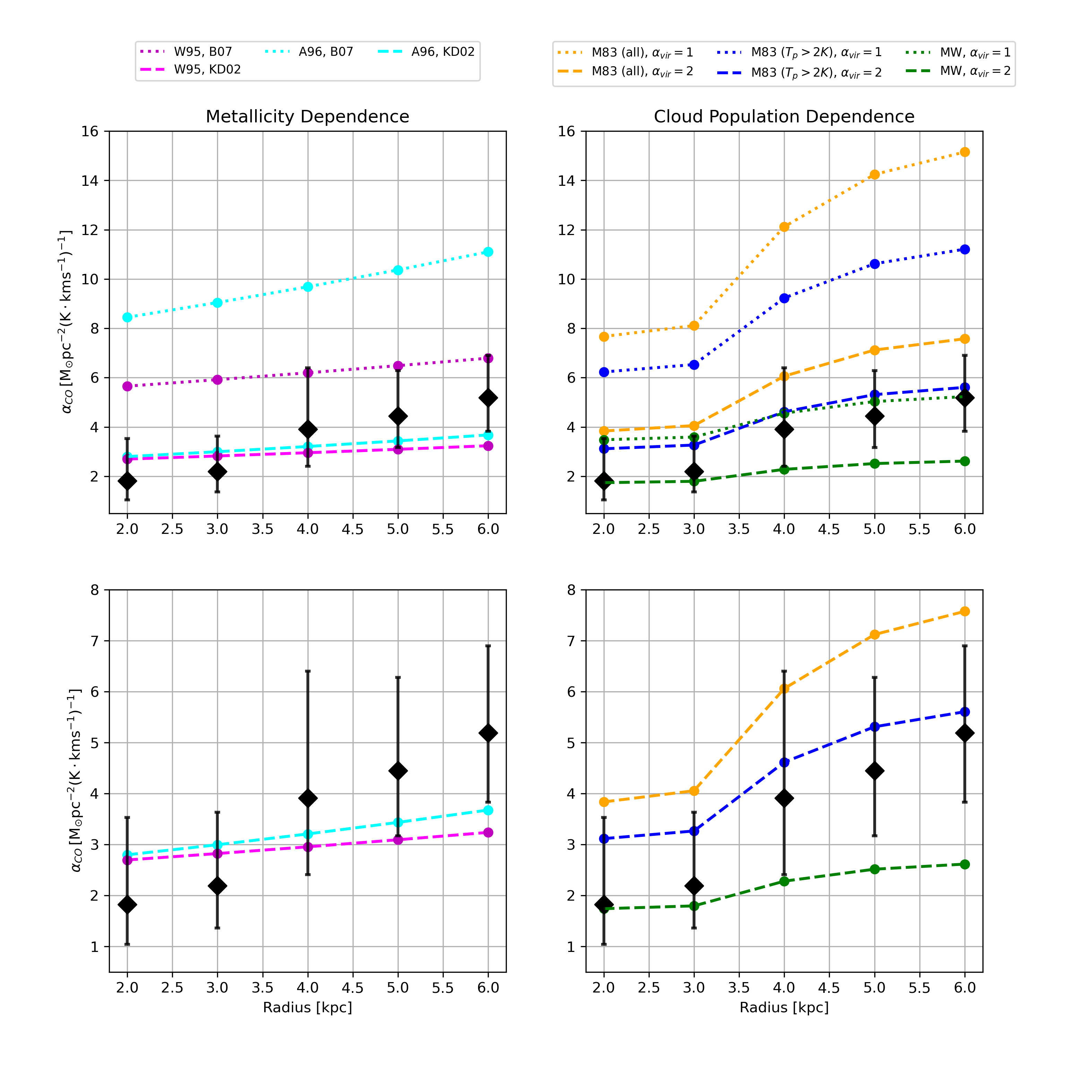}    \caption{Comparison of the $\alphaCO$ predicted by the metallicity dependent relations (left column) and by the cloud population-based model (right column) as labeled (see text), with the results in this study from the $\chi^2$ method (black diamonds). The results from the log $\Delta \GDR$ minimization method are similar to those derived from the $\chi^2$ method, and are omitted for clarity (see Figure \ref{fig:radial_var}). The bottom panels are the same as the top, but only for a smaller $\alphaCO$ range around our results, including the $\alphaCO$ predicted (bottom left) using the metallicity relations under the KD02 calibration, and (bottom right) if the clouds are marginally bound with $\alpha_{\rm vir} = 2$. The $\alphaCO$ predicted based on this particular calibration for the metallicity gradient or $\alpha_{\rm vir}$ = 2 for clouds are consistent with the $\alphaCO$ derived in this study, within uncertainty.}
    \label{fig:overplotavgacov1}
\end{figure*}

With these parameters, we derive the population-averaged $\langle \alpha_{CO} \rangle$ for clouds in M83.
The results in all cases are listed in Table \ref{tab:modelaco} (columns 6-11).
The $\langle \alpha_{CO} \rangle$ obtained in each annulus reproduces the radially increasing trend (Table \ref{tab:acovar}).
If we adopt the case of gravitationally-bound clouds ($\alpha_{\rm vir}=2$) as an example,
the $\langle \alpha_{CO} \rangle$ increases radially from $\alphaCO$ = 3.84 to 7.58 using ($\alpha_{\rm ref}$, $\beta$) obtained for all M83 clouds (M83$_{\rm all}$), $\alphaCO$ = 3.12 to 5.61 for those of the \textit{likely-resolved} clouds in M83 (M83$_{\rm >2K}$), and $\alphaCO$ = 1.74 to 2.62 for those of the MW clouds (MW), from the center ($R$ = 1-3 kpc) to the outer parts ($R$ = 5-7 kpc).
These increases are similar to the increase from $\alphaCO$ = 1.0-3.5 to 3.8-6.9 derived in this study (Table \ref{tab:newresults}).
Figure \ref{fig:overplotavgacov1} compares the $\alphaCO$ predicted by this cloud population-based model with the results in this study (Section \ref{sec:results}).

We should note that because $\langle \alpha_{CO} \rangle \propto 1/\alpha_{\rm vir}$ (eq. \ref{eqn:acoPL})
the absolute values increase by a factor of two if we adopt $\alpha_{\rm vir}=1$ (all clouds are in virial equilibrium; see Table \ref{tab:modelaco} columns 6-8), instead of $\alpha_{\rm vir}=2$ (the clouds are merely gravitationally-bound; columns 9-11).
This uncertainty stems from our lack of knowledge on the exact dynamical state of molecular clouds.

Here, we tested the idea that $\alphaCO$ may depend on the underlying cloud population and its variations,
and showed that the cursory estimation can explain the radial increase of $\alphaCO$.
There are of course some caveats.
For example, we did not consider a potential dependence of the cloud luminosity function (eq. \ref{eq:lumfunc}) on metallicity.
It is possible that previous studies might have already unintentionally treated the possible dependence on cloud population as a dependence on other parameters, since the variation of cloud population itself could depend on other parameters, e.g., dynamical environment, metallicity, etc. These dependencies make it difficult to isolate the origin of the $\alphaCO$ variations. Put simply, if the properties of molecular clouds vary, e.g., as in eq. (\ref{eq:viraco2}), for any reason,
the $\alphaCO$ averaged over a cloud population should also change accordingly.

\begin{deluxetable*}{ccc}[h]
\tablecaption{Parameters of the cloud luminosity function.}
\tablehead{
\colhead{(1)} & \colhead{(2)} & \colhead{(3)} \\ 
 \colhead{Radius} &  \colhead{$\gamma$} &  \colhead{$L_{{\rm max}}$}}
\startdata
\hline
1.0-3.0 & -1.67$^{-0.01}_{+0.01}$ & 2.7e6  \\ 
2.0-4.0 & -1.70$^{-0.01}_{+0.01}$ & 2.6e6  \\ 
3.0-5.0 & -1.88$^{-0.01}_{+0.01}$ & 1.3e6  \\ 
4.0-6.0 & -1.99$^{-0.01}_{+0.01}$ & 1.1e6  \\ 
5.0-7.0 & -2.03$^{-0.02}_{+0.01}$ & 9.7e5
\enddata
\tablecomments{ \\
(1) Radial range of aperture for fit in kpc. 
(2) $\gamma$ for the truncated power-law model. 
(3) L$_{{\rm{max}}}$ = $M_{{\rm{max}}}/\alpha_{{\rm{MW}}}$ in {$[{\rm K}  \kmps]^{-1} \cdot \pc^{2}$}. 
All parameters are obtained with the catalog of molecular clouds in M83 (Hirota et al., in prep.).
 These parameters are consistent with those in other studies of the inner Galaxy and other galactic disks similar to M83 \citep{%Rosolowsky:2005uf,
 Rosolowsky2005,
 Roman-Duval:2010fk, Colombo:2014aa}, although some studies reported steeper slopes typically in lower-density environments \citep{Rosolowsky2005, Colombo:2014aa}.}
\label{tab:clumfunc}
\end{deluxetable*}

\begin{deluxetable*}{cccccccccccccc}[h]
\tablecaption{Predicted $\alphaCO$ by the metallicity and cloud population dependences.\label{tab:acovar}}
\tablehead{
\colhead{(1)} & \colhead{(2)} & \colhead{(3)} & \colhead{} &\colhead{(4)} & \colhead{(5)} & \colhead{} & \colhead{(6)} & \colhead{(7)} & \colhead{(8)} & \colhead{} & \colhead{(9)} & \colhead{(10)} & \colhead{(11)}\\ 
\colhead{} & \multicolumn{2}{c}{B07 calibration} & \colhead{} & \multicolumn{2}{c}{KD02 calibration} & \colhead{} & \multicolumn{3}{c}{$\alpha_{\rm vir}=1$} & \colhead{} & \multicolumn{3}{c}{$\alpha_{\rm vir}=2$} \\ 
\cline{2-3} \cline{5-6} \cline{8-10} \cline{12-14} 
 \colhead{Radius} & \colhead{W95} & \colhead{A96} & \colhead{} & \colhead{W95} & \colhead{A96} && \colhead{M83$_{\rm all}$} & \colhead{M83$_{\rm >2K}$} & \colhead{MW} & \colhead{} & \colhead{M83$_{\rm all}$} & \colhead{M83$_{\rm >2K}$} & \colhead{MW} 
 }
\startdata
\hline
1.0-3.0 & 5.66 & 8.45  && 2.70 & 2.80 && 7.67  & 6.23  & 3.48  && 3.84 & 3.12 & 1.74 \\ 
2.0-4.0 & 5.93 & 9.05  && 2.82 & 3.00 && 8.11  & 6.53  & 3.59  && 4.06 & 3.27 & 1.80 \\ 
3.0-5.0 & 6.20 & 9.68  && 2.95  & 3.21 && 12.12 & 9.23  & 4.56  && 6.06 & 4.62 & 2.28 \\ 
4.0-6.0 & 6.49 & 10.37 && 3.09 &  3.43 && 14.24 & 10.62 & 5.03  && 7.12 & 5.31 & 2.52 \\ 
5.0-7.0 & 6.80 & 11.10 && 3.24 &  3.68 && 15.15 & 11.21 & 5.23  && 7.58 & 5.61 & 2.62 \\
\enddata
\tablecomments{ \\
(1) Radial range of aperture for fit in kpc.
(2-5) $\alphaCO$ predicted from metallicity dependence, using the M83 metallicity gradient of \citet{Bresolin:2009ce}.
The metallicity calibrations are based on ``B07"=\citet{Bresolin:2007aa} and ``KD02"=\citet{Kewley:2002aa}.
The $\alphaCO$-metallicity relations are taken from ``W95"=\citet{Wilson:1995lr} and ``A96"=\citet{Arimoto:1996aa}.
(6-11) $\alphaCO$ predicted from the cloud population (eq. \ref{eq:viraco2}), when $\alphaCO=1$ (virialized) and $\alphaCO=2$ (marginally-bound),
using the $\alpha_{\rm vir} \alphaCO$-$L_{\rm CO}$ relation (eq. \ref{eqn:acoPL}) of all clouds in M83 (``M83$_{\rm all}$") and (likely) resolved clouds in M83 (``M83$_{\rm >2K}$") in M83, and MW clouds (``MW"; see Table \ref{tab:arefmodels}).
The parameters for the cloud luminosity function (eq. \ref{eq:lumfunc}) are in Table \ref{tab:clumfunc}.} 
\label{tab:modelaco}
\end{deluxetable*}

\section{Summary and Conclusions}

We present new HI observations in the disk of M83 combining JVLA and single-dish GBT data. In the region we analyzed in this study, the new observations recover a factor of 2 more flux than the distributed THINGS moment 0 map used in previous studies.
We compare these HI observations to CO(J=1-0) and a dust surface density map derived 
with the conversion equation from a 500 $\mu$m image  \citep{Aniano:2020aa}. We derive ($\alphaCO$, $\GDR$) simultaneously with the dust-based approach, i.e., using dust masses as a tracer of total gas masses via two methods: (1) $\chi^2$ minimization (plane-fitting) and (2) log $\GDR$ minimization. These gave similar results. The dust-based approach requires accurate $\Sigma_{\rm{HI}}$ measurements as $\Sigma_{\rm HI}$ determines the absolute scale of $\GDR$, which calibrates the scale for $\alphaCO$. 

We find that $\alphaCO$ and $\GDR$ both increase as a function of radius, by a factor of $\sim$ 2-3 from R = 1-3 to 5-7~kpc. These radial changes are consistent with the ones found in previous studies, when the errors in flux calibration in the previous studies are taken into account.
As representative values for the galactic disk, we also average the annulus values and derive, e.g., luminosity-weighted averages of $\alphaCO$ = 3.14 (2.06, 4.96) and $\GDR$ = 137 (111, 182).
These are similar to the values measured in the Milky Way.

The variations of $\alphaCO$ and $\GDR$ in galaxies are often explained by metallicity dependence.
We show that the metallicity dependence could explain a factor of $1.2-1.3$
radial increase in $\alphaCO$.
 We also present another possibility: the radial variation of $\alphaCO$ as due to the variation of the underlying cloud population. This idea stems from the observed nonlinearity between $\Sigma_{\rm{d}}$ and $\Sigma_{\rm{H}_2}$ 
(assuming it primarily arises from gas properties, and not potential systematic uncertainties in dust),
and the tracing of different ranges of $I_{\rm{CO}}$ in each annulus. We test this possibility by assuming constant $\alpha_{\rm vir}$ for clouds, and using a 
truncated power-law cloud luminosity function and an empirical dependence of $\alphaCO$ on cloud CO luminosity. We show that the cloud averaged $\langle \alphaCO \rangle$ can explain a factor of $1.5-2.0$ change in $\alphaCO$, compared to the factor of $2-3$ change 
from the annulus-based fits. 
Beyond the relative radial changes, the absolute values of $\alphaCO$ predicted by the metallicity and cloud population-dependences are consistent with the results we obtained, if we adopt a particular calibration for the metallicity measurements and $\alpha_{\rm vir}$ value (Figure \ref{fig:overplotavgacov1}). 
 This is suggestive, but must be revisited in the future, as there is no strong evidence to support the particular calibration nor value.

The exact origin of $\alphaCO$ variations in nearby galaxies remains difficult to disentangle as most environmental parameters vary with galactic radius and correlate with each other. 
While the dependence on cloud population 
may in fact embed dependencies on dynamical environment, disk potential, metallicity, etc., it is important to recognize that $\alphaCO$ could depend on cloud population, 
especially as we move towards more spatially resolved analyses of molecular clouds. 
Similarly, in terms of cloud population and their properties, we discussed only about the dependence of $\alphaCO$ on cloud luminosity. 
$\alphaCO$ is also known to vary with cloud temperature $T$ and density $\rho$ (i.e., $\alphaCO \propto \rho^{1/2}/T$, ee Appendix \ref{sec:Dickman}) \citep{Dickman:1986aa, Solomon:1987pr, Heyer:2001aa}. 
These parameters could also depend on each other.

\begin{appendix}

\onecolumngrid
\section{The Conversion Factor and Virial Parameter of Molecular Clouds in M83 \label{sec:m83fits}}

For the discussion in Section \ref{sec:clouddependence},
we obtain the relationship among the conversion factor ($\alphaCO$), virial parameter ($\alpha_{\rm vir}$),
and CO luminosity ($L_{\rm CO}$) of molecular clouds in M83,
using a new cloud catalog (Hirota et al. in prep.).
Detailed discussions on this catalog and its limitations (e.g., cloud identification, measurements, and possible biases) are given in Hirota et al. in prep..
In particular, the clouds at low $L_{\rm CO}$ are unlikely to be resolved in their observations at a 40~pc resolution, since their counterparts in the MW are typically smaller.

\begin{figure*}[h]
    \centering
    \includegraphics[width=0.6\textwidth]{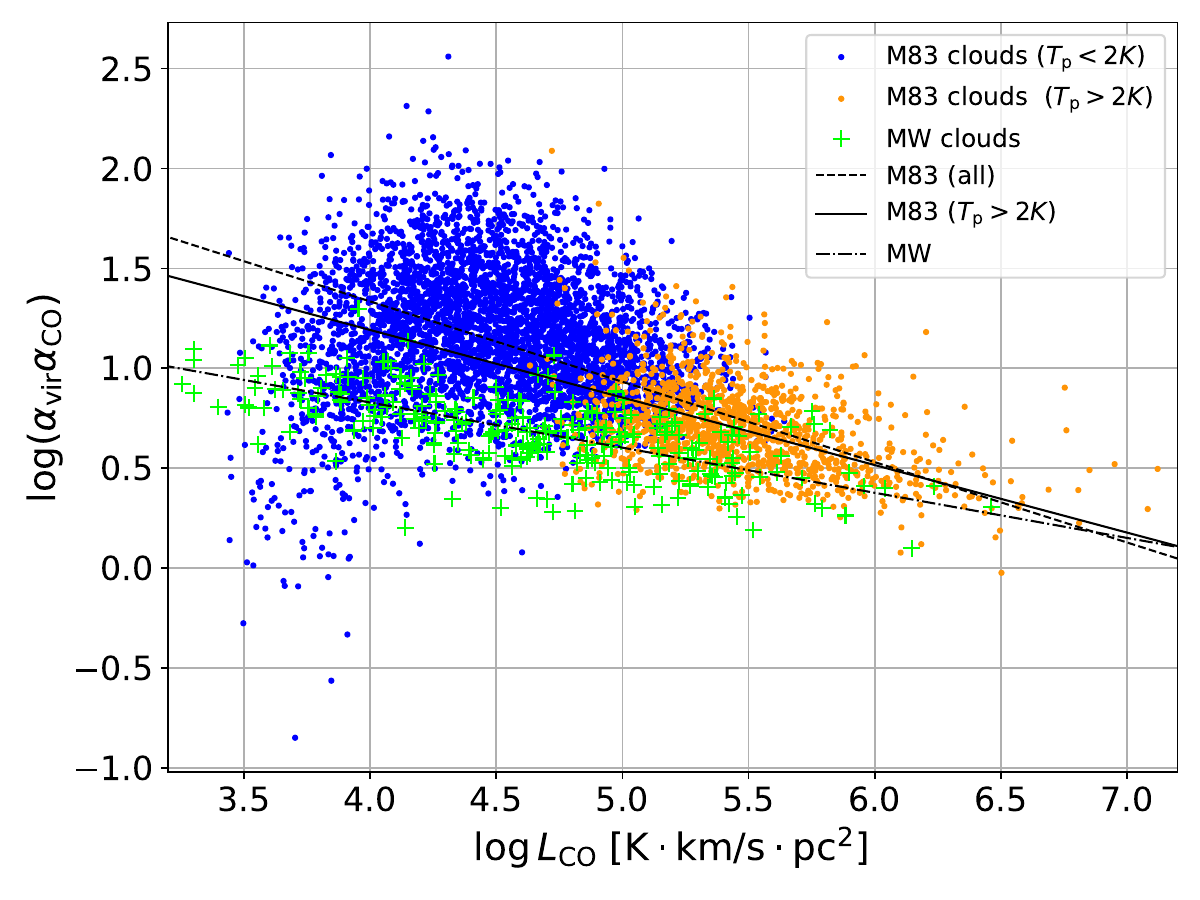}
    \caption{Comparison of $\alpha_{\rm vir} \alphaCO$ and $L_{\rm CO}$ of individual molecular clouds in M83 (Hirota et al. in prep.) and in the MW \citep{Solomon:1987pr}.
    The linear fits are made with the orthogonal distance regression, for all M83 clouds (dashed line), M83 clouds with $T_{\rm p}>2$~K (solid), and MW clouds (dot-dashed line).
    The peak brightness temperature $T_{\rm peak}$ is a crude
    measure to separate resolved clouds ($T_{\rm peak}>2$~K)  from unresolved ones  ($<2$~K), but this separation is enough for the purpose of the study presented in the main text.
    These fits may be subject to further refinement, but are sufficient for the discussion in Section \ref{sec:clouddependence}.
    Hirota et al. (in prep) will present more accurate results with possible biases taken into account.}
    \label{fig:hirota}
\end{figure*}

Figure \ref{fig:hirota} shows a correlation between the $\alpha_{\rm vir} \alphaCO$ and $L_{\rm CO}$ of individual molecular clouds in M83.
We adopt $\alpha_{\rm vir} \alphaCO= M_{\rm vir}/L_{\rm CO} = 9/2  (R \sigma_{\rm v}^2/G)/L_{\rm CO}$, where $R$ and $\sigma_{\rm v}$ are the radius and velocity dispersion of the cloud and $G$ is the gravitational constant.
For comparison, we also plot the clouds in the MW \citep{Solomon:1987pr}.

The lines are the fits to all M83 clouds (dashed line), M83 clouds with a peak temperature of $T_{\rm peak}>2$~K (solid), and MW clouds (dot-dashed line).
Hirota et al. (in prep.) uses the peak brightness temperature $T_{\rm peak}$ as a rough measure to separate likely-resolved clouds ($T_{\rm peak}>2$~K) from likely-unresolved ones  ($<2$~K), although the separation is hardly perfect. They adopted $T_{\rm peak}$ for this separation since it is among the direct observables in their data cube. Cloud radius may be an alternative parameter for this separation; however, while it can be measured, it is uncertain for clouds with sizes close to or smaller than the beam size.
Galactic clouds appear to have $T_{\rm peak}>2$~K when their surface brightness is averaged over entire cloud areas \citep[e.g., the Taurus molecular cloud; see ][]{Goldsmith:2008aa}. Hence, the clouds with $T_{\rm peak}<2$ K at the 40~pc resolution are likely suffering from a beam dilution effect (i.e., likely-unresolved). In addition, 
the majority of the clouds in Hirota et al. in prep. 
with $T_{\rm peak}>2$~K have masses of 
$\gtrsim 2\times 10^5 \Msun$
when they are calculated from $L_{\rm CO}$, assuming the Galactic $\alphaCO$. 
Using the scaling relations by \citet{Solomon:1987pr}, these masses correspond to cloud diameters of $\sim$40~pc, which are about their resolution.
The clouds with $T_{\rm peak}<2$~K are typically less massive and are likely smaller.  
More detailed discussions are in Hirota et al. (in prep.).

We adopt orthogonal distance regression (ODR) for the fitting.
The slopes and $y$-intercepts of the lines are listed in Table \ref{tab:arefmodels}.
We note that Hirota et al. in prep. will derive and discuss
these relations more carefully by taking into account possible biases in the catalog.

\begin{deluxetable*}{ccccc}[h]
\tablecaption{Fit results between $\log \alpha_{\rm vir} \alphaCO$ vs. $\log L_{\rm CO}$ (eq. \ref{eqn:acoPL}) and corresponding $\alpha_{\rm ref}$ assuming $L_{\rm ref}$ = 10$^5$ (eq. \ref{eq:viraco2}). 
\label{tab:arefmodels}}
\tablehead{
\colhead{Clouds} & \colhead{Slope ($\beta$)} & \colhead{$y$-intercept} & \colhead{$\alpha_{\rm ref}$}}
\startdata
\hline
M83 clouds (all)             &  -0.40 & 2.94 & 8.56\\ 
M83 clouds ($T_{\rm p}>2$~K) &  -0.34 & 2.54 & 7.14\\
MW clouds                    &  -0.23 & 1.73 & 4.01
\enddata
\end{deluxetable*}

\onecolumngrid
\section{Consideration of Metallicity Dependence} \label{sec:metallicity}

\begin{figure*}[h]
    \centering
    \includegraphics[width=1.0\textwidth]{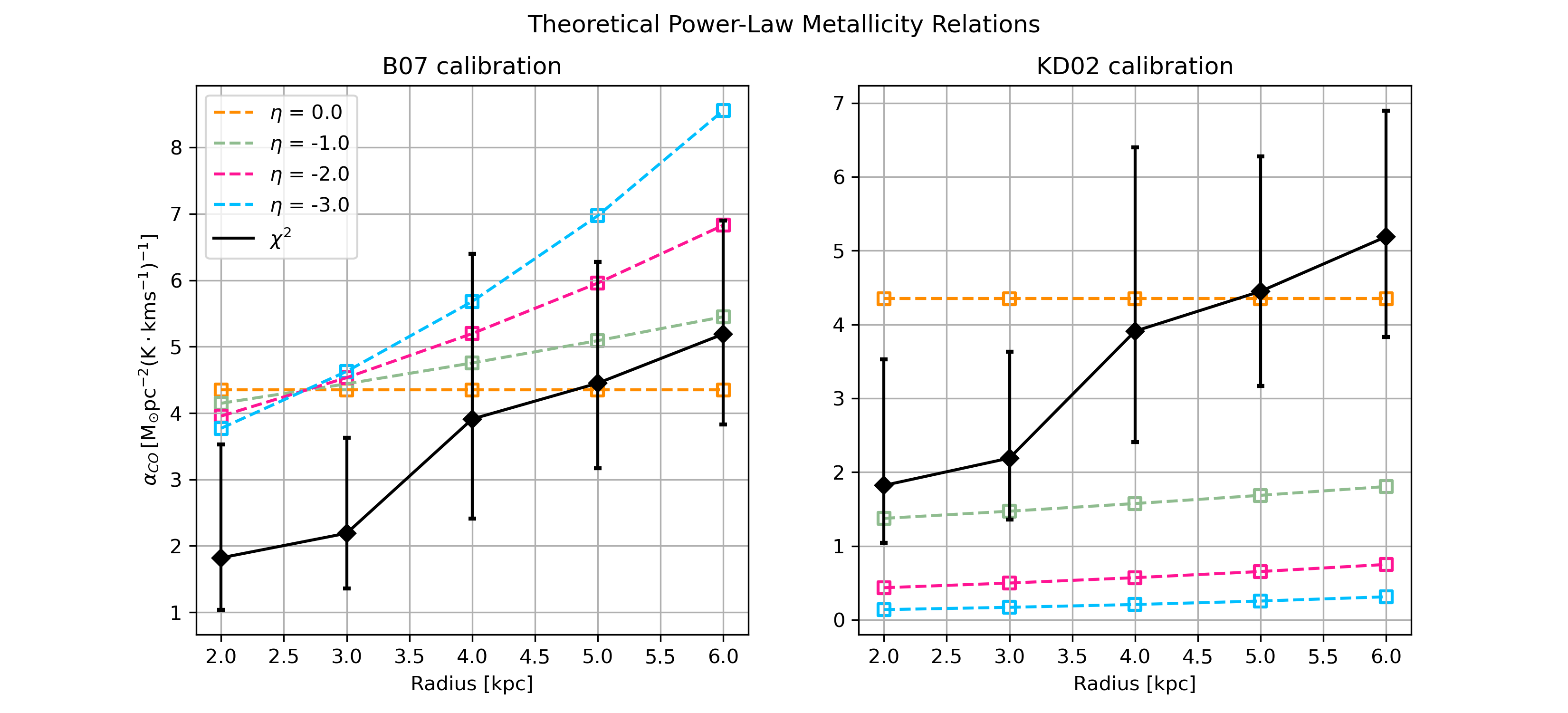}
    \caption{$\alphaCO$ in M83 as predicted by adopting a power-law metallicity relation for a range of indices $\eta$ (orange, green, pink, and blue) suggested by observational studies in the literature, and the results in this study from the $\chi^2$ method (black diamonds). The left and right panels use the metallicity calibrations from \citep[][B07]{Bresolin:2007aa} and \citep[][KD02]{Kewley:2002aa}.}
    \label{fig:metalPL}
\end{figure*}

Variations in $\alphaCO$ are often attributed to a dependence on metallicity. 
Recent observational studies have suggested $\alphaCO$ prescriptions, i.e., $\alphaCO$ $\propto \left(\frac{Z}{Z_{\odot}}\right)^{\eta}$, with a range of power-law indices \citep[$\eta$ = -0.7 to -3.4; ][]{Schruba:2012aa, Hunt:2015aa, Accurso2017, Madden:2020aa} . 
While there are not many alternative ways to constrain $\alphaCO$ especially in low-metallicity environments, many of these studies adopt assumptions that could be questioned (e.g., the universality of the star formation process/Kennicutt-Schmidt relation independent of environments, the accuracy of chemical models that assume geometry, density, and temperature of the gas).
There are also theoretical works that analyzed simulations to understand the metallicity-dependence of $\alphaCO$ \citep{Gong:2020aa, Hu2022},
which reproduced the observed $\alphaCO$-metallicity dependences with power-law indices of -0.80 and -0.70.
Hence, instead of adopting a single relation directly, we explore a range of indices encompassing suggested $\alphaCO-Z$ relations in the literature. For this, we adopt the power-law model $\alphaCO$ = $\alpha_{{\rm{MW}}}\left(\frac{Z}{Z_{\odot}}\right)^{\eta}$ with $\eta = 0.0, -1.0, -2.0$ and $-3.0$. We use the metallicity gradient for M83 and calibrations described in section \ref{sec:metaldependence}. In Table \ref{tab:powermetal}, we show the results from this model as a function of radius. In Figure \ref{fig:metalPL}, we compare them with the $\chi^2$ results obtained in section \ref{sec:results}. The absolute value of the predicted $\alphaCO$ depends on the metallicity calibration adopted, but $\eta=$ $-2.0$ or $-3.0$ (the steepest slopes among the suggested) with the B07 calibration, may potentially explain the factor of $\sim$2-3 radial variation in $\alphaCO$ derived in section \ref{sec:results}. For $\eta=$ $-2.0$ or $-3.0$ with the B07 calibration to match in absolute value with the $\chi^2$ results, $\alpha_{{\rm{MW}}}$ would have to be $\sim$2 times smaller.

\begin{deluxetable*}{cccccccccccc}[h]
\tablecaption{Predicted $\alphaCO$ when adopting a power-law metallicity dependence, i.e., $\alphaCO = \alpha_{{\rm{MW}}}\left(\frac{Z}{Z_{\odot}}\right)^\eta$.}\label{tab:powermetal}
\tablehead{\colhead{(1)} & \colhead{(2)} & \colhead{(3)} & \colhead{(4)} &\colhead{(5)} & \colhead{} & \colhead{(6)} & \colhead{(7)} & \colhead{(8)} & \colhead{(9)} \\ 
\colhead{} & \multicolumn{4}{c}{B07 calibration} & \colhead{} & \multicolumn{4}{c}{KD02 calibration} \\ 
\cline{2-5} \cline{7-10} 
 \colhead{Radius} &\colhead{0.0} & \colhead{-1.0} & \colhead{-2.0} & \colhead{-3.0} & \colhead{} & \colhead{0.0} & \colhead{-1.0}  & \colhead{-2.0} & \colhead{-3.0}}
\startdata
\hline
1.0-3.0 & 4.35 & 4.15  & 3.95 & 3.77 && 4.35 & 1.37 & 0.43 & 0.14 \\ 
2.0-4.0 & 4.35 & 4.44  & 4.53 & 4.63 && 4.35 & 1.47 & 0.50 & 0.17 \\
3.0-5.0 & 4.35 & 4.75  & 5.20 & 5.68 && 4.35 & 1.57 & 0.57 & 0.21 \\ 
4.0-6.0 & 4.35 & 5.09  & 5.96 & 6.97 && 4.35 & 1.69 & 0.65 & 0.25 \\ 
5.0-7.0 & 4.35 & 5.45  & 6.83 & 8.56 && 4.35 & 1.80 & 0.75 & 0.31 \\
\enddata
\tablecomments{ \\
(1) Radial range of aperture in kpc.
(2-9) $\alphaCO$ predicted with the power-law metallicity dependence, for {$\eta$ = 0, -1.0, -2.0, and -3.0}, using the M83 metallicity gradient of \citet{Bresolin:2009ce}. We adopt $Z/Z_{\odot}$ = 1 at 12+log(O/H) = 8.69 \citep{Asplund:2009aa}.
 The metallicity calibrations are based on ``B07"=\citet{Bresolin:2007aa} and ``KD02"=\citet{Kewley:2002aa}.}
\end{deluxetable*}

\onecolumngrid
\section{Dickman et al. 1986}\label{sec:Dickman}

\citet{Dickman:1986aa} calculated $\alphaCO$ as an average over the cloud population across the disk of the MW. In essence, they adopted [1] one empirical relation, the size-velocity dispersion relation $\sigma_v \propto R^{\gamma}$, and assumed [2] virial equilibrium $\alpha_{\rm vir}=1$ (hence,  $M_{\rm{vir}} \propto R \sigma_{v}^2$), [3] $L_{\rm{CO}} \propto T\sigma_{v}R^{2}$, and [4] a constant surface brightness temperature $T$ for all clouds. From these, they derived
\begin{eqnarray}
    \alphaCO = \frac{M_{\rm{vir}}}{L_{\rm{CO}}} \propto \frac{1}{T}R^{\gamma - 1},  \label{eq:Dickman}
\end{eqnarray}
i.e., a dependence of $\alphaCO$ on cloud size (radius) under constant $T$. With this $\alphaCO$-$R$ relation, they averaged $\alphaCO$ over the distribution of $R$ in the MW cloud population, and obtained $\alphaCO$ $\sim$ 4. 
We extended this idea, applied it to varying cloud populations, and explained the variations of $\alphaCO$ with galactic radius in M83.
We used [1] an empirical relation of eq. (\ref{eq:viraco2}) under the assumptions equivalent to [2] and [3] -- in our case, we set $\alpha_{\rm vir}$=1 or 2 and used $L_{\rm{CO}}$ measured by integrating $T$ over a cloud area (equivalent to the $R^2$ term) and velocity ($\sigma_{v}$).
By starting from eq. (\ref{eq:viraco2}) instead of eq. (\ref{eq:Dickman}), we did not need to adopt the assumption [4] of constant $T$.

As a further note, from assumptions [1]-[3], the volume density of cloud is $\rho \propto M_{\rm{vir}}/R^3 \propto R^{2(\gamma -1)}$. With eq. (\ref{eq:Dickman}), we can derive
\begin{equation}
    \alphaCO \propto \frac{\rho^{1/2}}{T}
\end{equation}
independent of $\gamma$.

\end{appendix}

\clearpage

\begin{acknowledgments}
We thank Bruce Draine for explanations on the dust model,
Tony Graham and Nick Ferraro at NRAO Helpdesk for their advice on JVLA data reduction, and 
Ananthan Karunakaran for generously allowing us to take calibration data at GBT in their allocated time.
We also thank the anonymous referee for suggestions.
JK acknowledges support from NSF through grants AST-1812847 and AST-2006600.
FE is supported by JSPS KAKENHI grant No. 17K14259 and 20H00172.
The National Radio Astronomy Observatory is a facility of the National Science Foundation operated under cooperative agreement by Associated Universities, Inc.
The Green Bank Observatory is a facility of the National Science Foundation operated under cooperative agreement by Associated Universities, Inc.
This paper makes use of the ALMA data, ADS/JAO.ALMA\#2017.1.00079.S.
ALMA is a partnership of ESO (representing its member states), NSF (USA) and NINS (Japan), together with NRC (Canada), MOST and ASIAA (Taiwan), and KASI (Republic of Korea), in cooperation with the Republic of Chile. The Joint ALMA Observatory is operated by ESO, AUI/NRAO and NAOJ.
Herschel is an ESA space observatory with science instruments provided by European-led Principal Investigator consortia and with important participation from NASA.
This research has made use of the NASA/IPAC Extragalactic Database (NED), which is operated by the Jet Propulsion Laboratory, California Institute of Technology, under contract with the National Aeronautics and Space Administration.
\end{acknowledgments}

\facilities{JVLA, GBT, ALMA, Herschel, IRSA, NED}

\bibliography{ms.bbl}

\end{document}